# Robust Survivability-Oriented Scheduling of Separable Mobile Energy Storage and Demand Response for Isolated Distribution Systems

Wei Wang, *Member, IEEE*, Yufei He, Xiaofu Xiong, *Member, IEEE*, and Hongzhou Chen.

***Abstract*—Extreme circumstances in which a local distribution system is electrically isolated from the main power supply may not always be avoidable. Efforts must be made to keep the lights on for such an isolated distribution system (IDS) until reconnection to the main power source. In this paper, we propose a strategy to enhance IDS survivability by utilizing the coordination of two flexible approaches, namely, separable mobile energy storage systems (SMESSs), which construct non-wires links for energy transmission between the IDS and the external live power sources, and demand response (DR), which adjusts the internal electrical demand of the IDS to provide effective operating stress alleviation. Considering the uncertainty of renewable energy generation and loads, a two-stage robust optimization (RO) model involving the joint scheduling of these two approaches is constructed. The objective is to minimize the fuel consumption and the decreased and unserved demand under the worst-case scenario to endow the IDS with extended survivability. Finally, test is conducted and the results demonstrate the effectiveness of the proposed method in enhancing the survivability of IDS.**

***Index Terms*—Isolated distribution system, separable mobile energy storage systems, demand response, robust optimization.**

## NOMENCLATURE

*Sets*

| | |
|---|---|
| $\mathcal{T}$ | Set of time spans of scheduling. $\mathcal{T}=\{t\|1\le t\le D, t\in\mathcal{Z}\}$, where $D=\|\mathcal{T}\|$ and $\mathcal{Z}$ is set of integers. |
| $\mathcal{N}$, $\mathcal{N}_{DR}$ | Set of nodes in the IDS and those participating in DR. |
| $\mathcal{S}$, $\mathcal{S}_S$, $\mathcal{S}_R$ | Sets of external sources, such as a substation, and those as REG isolated from the IDS, respectively. $\mathcal{S}=\mathcal{S}_S\cup\mathcal{S}_R$. |
| $\mathcal{N}_S$ | Set of nodes that support the access of SMESSs, including those in the IDS and the external sources. |
| $\mathcal{F}(i)$ | Set of FFGs located at node $i$ in the IDS. |
| $\mathcal{M}$, $\mathcal{K}$ | Sets of Carrs and Mods of SMESSs. |
| $\mathcal{L}$ | Set of branches in the IDS. |

*Variables*

| | |
|---|---|
| $\Psi_{11}$ | Fuel consumption for SMESSs' movement. |
| $\Psi_{12}$ | Fuel consumption of FFGs. |
| $\Psi_2$ | Total weighted energy demand reduced by DR during the scheduling. |
| $\Psi_3$ | Total weighted nonpicked-up energy demand during the scheduling. |
| $\chi_i$ | Sum of the nonpicked-up power load at node $i$. |
| $x_{j,i,t}$ | Binary variable, 1 if Carr $j$ is parked at node $i$ during time span $t$, 0 otherwise. |
| $v_{j,i,t}$ | Binary variable, 1 if Carr $j$ is traveling to node $i$ during time span $t$, 0 otherwise. |
| $S_{j,t}$ | Travel time to be consumed by Carr $j$ during time span $t$. |
| $R_{j,t}$ | Residual travel time of Carr $j$ during time span $t$. |
| $\omega_{j,t}$ | Binary variable, 1 if Carr $j$ is traveling during time spans $t-1$ and $t$. |
| $\zeta_{k,i,t}$ | Binary variable, 1 if Mod $k$ belongs to node $i$ during time span $t$, 0 otherwise. |
| $\gamma_{k,j,t}$ | Binary variable, 1 if Mod $k$ belongs to Carr $j$ during time span $t$, 0 otherwise. |
| $\alpha_{j,i,k,t}$ | Binary variable, 1 if Carr $j$ carrying Mod $k$ arrives at node $i$ during time span $t$, 0 otherwise. |
| $c_{k,i,t}/d_{k,i,t}$ | Binary variable, 1 if Mod $k$ is charged/discharged at node $i$ during time span $t$, 0 otherwise. |
| $P^{c.S}_{k,i,t}/P^{d.S}_{k,i,t}$ | Active power output of Mod $k$ charged/discharged at node $i$ during time span $t$. |
| $Q^{S}_{k,i,t}$ | Reactive power output of Mod $k$ charged/discharged at node $i$ during time span $t$. |
| $SOC_{k,t}$ | State of charge of Mod $k$ at the end of time span $t$. |
| $\delta_i$ | Binary variable, 1 if the load at node $i$ is picked up in the scheduling, 0 otherwise. |
| $\rho_{i,t}$ | Binary variable, 1 if DR is executed at node $i$ during time span $t$, 0 otherwise. |
| $P^{DR}_{i,t}$ | Active load reduction due to DR execution at node $i$ during time span $t$. |
| $C_{i,t}$ | Accumulated energy (within a single DR execution) during time span $t$ at node $i$ |
| $EP_{i,t}$ | Energy about to rebound at node $i$ after time span $t$. |
| $P^{EP}_{i,t}$ | Net rebounded active load due to energy payback at node $i$ during time span $t$. |

$P^{\mathrm{F}}_{f,t}$/ $Q^{\mathrm{F}}_{f,t}$ Active/reactive power output of FFG $f$ during time span $t$.

$P^{\mathrm{L}}_{i,t}$/ $Q^{\mathrm{L}}_{i,t}$ Active/reactive power load at node $i$ due to scheduling during time span $t$.

$\tilde{P}^{\mathrm{OL}}_{i,t}$ Original active power load at node $i$ if it is picked up without DR during time span $t$.

$P^{\mathrm{IN.S}}_{i,t}$ / $Q^{\mathrm{IN.S}}_{i,t}$ Active/reactive power input from SMESSs at node $i$ during time span $t$.

$\tilde{P}^{\mathrm{REG}}_{i,t}$ Active power output at REG $i$ out of the IDS during time span $t$.

$P^{\mathrm{IN.F}}_{i,t}$ / $Q^{\mathrm{IN.F}}_{i,t}$ Active/reactive power input from FFGs at node $i$ during time span $t$.

$P_{i'i,t}$/ $Q_{i'i,t}$ Active/reactive power flow on branch ($i'$, $i$) from node $i'$ to node $i$ during time span $t$.

$V^2_{i,t}$ Squared voltage magnitude at node $i$ during time span $t$.

$\hat{u}^{\mathrm{L}}_{i,t}$/ $\check{u}^{\mathrm{L}}_{i,t}$ Upward/downward fluctuation of the original load at node $i$ during time span $t$.

$\hat{u}^{\mathrm{REG}}_{i,t}$ / $\check{u}^{\mathrm{REG}}_{i,t}$ Upward/downward fluctuation of the power output of REG $i$ during time span $t$.

*Parameters*

$\Psi_{1,\max}$, $\Psi_{2,\max}$, $\Psi_{3,\max}$ Estimated possible upper bounds of total fuel consumption, energy demand reduced by DR and nonpicked-up energy demand, respectively.

$\kappa_1$, $\kappa_2$, $\kappa_3$ Weight coefficients for $\Psi_1$, $\Psi_2$, $\Psi_3$.

$w_i$ Priority weight of the load at node $i$.

$\mu_j$, $\sigma_f$ Unit fuel consumption for travel of Carr $j$ and generation of FFG $f$.

$\bar{\psi}_1$ Remaining fuel at the start of the scheduling.

$\Delta t$ Length of a single time span.

$M$ A sufficiently large/small positive number.

$T_{j,ii'}$ Time spans spent traveling from node $i$ to node $i'$ for Carr $j$.

$W_k$ Capacity consumed by Mod $k$.

$A_j$ Carrying capacity Carr $j$.

$P^{\mathrm{c.S}}_{k,\max}$/$P^{\mathrm{d.S}}_{k,\max}$, $S_{k,\mathrm{Mod}}$ Maximum charging/discharging power and rated apparent power of Mod $k$.

$E_k$ Energy capacity of Mod $k$.

$e^c_k$/$e^d_k$ Charging/discharging efficiency of Mod $k$.

$SOC_{k,\min}$/ $SOC_{k,\max}$ Minimum/maximum allowable range of state of charge for Mod $k$.

$\tau^{\mathrm{DR}}_{i,\min}$/ $\tau^{\mathrm{DR}}_{i,\max}$ Lower/upper bound of the ratio of load reduction due to DR execution at node $i$.

$\eta_i$ Ratio of the reactive load to the active load at node $i$.

$\rho'_{i,t}$ Record of whether DR was executed at node $i$ or not during time span $t$ in the previous scheduling.

$T_{i,\mathrm{DU,max}}$/ $T_{i,\mathrm{DU,min}}$ Maximum/minimum allowable duration of a single DR execution at node $i$.

$T_{i,\mathrm{IN,min}}$ Minimum allowable interval between two adjacent DR executions at node $i$.

$T_{i,\mathrm{DR,max}}$ Maximum allowable total duration of DR executions at node $i$ in the scheduling.

$C'_{i,t}$, $EP'_{i,t}$ Records of the accumulated energy and the energy about to rebound at node $i$ during time span $t$ in the previous scheduling.

$T_{i,\mathrm{pdu}}$ Duration of an energy payback at node $i$.

$b_{i,h}$ Gain coefficient of energy payback on the active load during the $h$th time span in an energy payback.

$P^{\mathrm{F}}_{f,\max}$/ $Q^{\mathrm{F}}_{f,\max}$, $S_{f,\mathrm{FFG}}$ Maximum active/reactive power output and rated apparent power of FFG $f$.

$P^{\mathrm{sub}}_{i,\max}$ Maximum spare power at substation $i$ out of the IDS.

$r_{ii'}$/ $x_{ii'}$ Resistance/reactance of branch ($i$, $i'$).

$V_{i,\min}$/ $V_{i,\max}$ Lower/upper bound of the voltage magnitude at node $i$.

$S_{i'i,max}$ Apparent power capacity of branch ($i'$, $i$).

$\bar{P}^{\mathrm{OL}}_{i,t}$ , $\bar{P}^{\mathrm{REG}}_{i,t}$ Forecasted active power load and REG output at node $i$ during time span $t$.

$\Delta\hat{P}^{\mathrm{OL}}_{i,t}$ /$\Delta\check{P}^{\mathrm{OL}}_{i,t}$ Maximum upward/downward fluctuation of the original load at node $i$.

$\Delta\hat{P}^{\mathrm{REG}}_{i,t}$ / $\Delta\check{P}^{\mathrm{REG}}_{i,t}$ Maximum upward/downward fluctuation of the power output of REG $i$.

# I. Introduction

CATASTROPHIC events over the past decades and the increasing reliance of society on electricity have raised awareness of the urgent demand and significance for enhancing power system resilience under high-impact, low-frequency (HILF) events. A resilient power system, according to EPRI reports [1], [2], should 1) be hardened to limit damage, 2) quickly restore the electric service, and 3) aid customers in continuing some level of service without access to normal power sources, referring to the three elements of resilience: *prevention*, *recovery*, and *survivability*, respectively.

In contrast to other parts of the power system, the distribution system (DS)'s greater exposure, complexity, and geographic reach result in higher vulnerability to most kinds of disruptions particularly HILF events that could cause widespread and long-term outages [1], [2]. To enhance DS resilience, extensive studies have been conducted on the first two elements. Regarding *prevention*, researchers have mainly focused on planning and reinforcement of facilities; and measures relating to line hardening, the allocation of energy resources such as energy storage and distributed generation [3], automatic switch installation [4], and proactive islanding [5] have been studied. Among the research regarding *recovery*, which aims at restoring electric service of DS rapidly after the onset of HILF events, in addition to the pre-allocated energy resources that can work soon to restore the power supply, measures involving the scheduling of mobile energy resources (*e.g.*, mobile energy storage systems (MESSs), mobile generators) [6], [7], repair crew, microgrids formation by DS reconfiguration [8], and demand response (DR) [9] have been shown to be effective. Enhanced situational awareness and precise damage assessment also make a large difference in DS recovery [10].

While current research mainly focuses on enhancing DS resilience from the aspects of *prevention* and *recovery*, fewer studies have addressed the concerns regarding the other aspect *survivability*. The basic function of power system, especially of the DS, is known to provide customers with continuous power supply. A "living" DS means this function is still active without being completely lost; and that a DS survives a major disaster can be intuitively reflected as the sustentation of power supply to some, even if not all, customers it takes charge of. As stated by EPRI, *survivability* focuses on the issue about how the electric service to customers is sustained when the local DS is isolated from its major power sources [1], [2], *e.g.*, the bulk grid. Such isolation can happen suddenly, giving the DS little time to prepare for it in advance. As the fatal consequence, enormous power shortage, which may even last for long under some circumstances (*e.g.*, if the repair of faulted components takes days), will force the survival of the DS to be a top priority. Similar scenarios characterized by the isolated local power system are also highlighted in the definitions of survivability in other relevant studies. For example, in [13], survivability of microgrid is referred to as the probability that critical load will be fully served over the duration of an islanding event. Ref. [14] defines survivability of microgrid in islanded mode as the ability of the microgrid to withstand the loss of any of inner generators without collapsing or shedding of loads. In [15], survivability is stated as minimizing the load shed during the post-disturbance islanding operation of microgrid. In this paper, we mainly follow the definition from EPRI. Rather than aiming at restoring the lost loads rapidly as *recovery* does, in our opinion, *survivability* emphasizes the performance of an isolated DS (IDS) in sustaining the power supply to as many loads as possible over a longer duration until that isolation disappears, *i.e.*, until the IDS is reconnected to its main sources like the grid.

In this regard, most relevant studies have shown the efficacy of renewable energy generation (REG) and energy storage in boosting the survivability of whether individual customers that lose the grid-supplied power or islanded local systems such as microgrids. For example, a group from the University of Washington made a series of field trips to a remote town in Puerto Rico after Hurricane Maria heavily hit it in 2017 and they installed photovoltaic-battery systems for the local residents that would otherwise lose power for long. The collected data and further research show that the systems can support the electric power demand of users over a month and that photovoltaic-battery systems, which has high capital cost but low ongoing cost, will be more suitable to sustain the power to a place likely to experience months of outages [11]. In [12], flywheel energy storage systems, along with generators, are used as backup power sources of a data center. The flywheel systems can provide a short-term power support to the critical loads of the data center after a utility outage, and then the generators start up to supply the full loads and charge the flywheels. The research in [13] reveals a hybrid configuration, including energy storage, REG and diesel generators, can endow the microgrid with lower fuel consumption and higher survivability even after days of operation, in contrast to the one with only diesel generators. It is also demonstrated in [16] that the combination of batteries and REG helps to extend the time that a microgrid can survive a blackout by hours. Thus, from these studies, two points of keeping the lights on for long for the de-energized customers or DSs can be recognized: one is the sustainable supply of power and energy, and another is the compensation, or necessary reserve, to the mismatch between the supply and the demand to be served. Two effective tools are then introduced to cope with these two points for the IDS.

In addition to a proactive island formed pre-emptively ahead of an HILF event, an IDS can be mainly formed by the forced outage of lines that link the DS and its major power source, *e.g.*, the substation, the REG that powers a geographically isolated area like a remote island. Under this circumstance, the lifeline for electricity fed by the external source to the IDS is blocked until the out-of-service lines that link them get repaired and resume running. What is worse, the available capacity of power and energy resources within the IDS is not always sufficient or even none to serve the full demand, especially now that the distributed generation has not yet been penetrated intensively in the DSs. Seeing this through the eyes of the IDS operator, we may feel anxious not only because of the predicament faced by the IDS in terms of very insufficient supply, but also because the "stranded" source with spare power may be just a step away from the IDS but cannot be accessible. Encouragingly, the emergent MESS technology provides an effective solution to that due to its remarkable spatial flexibility.

As the most common function, energy storage typically fixed at a site can realize a temporal energy shift, *e.g.*, to level the REG power output or achieve arbitrage. While holding this ability, the MESS, through a simple modification on equipment, can further realize the energy shift in the spatial dimension: MESS can absorb energy from a place (*e.g.*, a feeder or source with spare power capacity) and, by moving, release it at another place in need of energy, even if the two places are electrically isolated. To put it another way, while electricity is known to be transmitted commonly by power lines, MESS provides another way for electricity transmission by roads or railways. Due to this, MESS has received increasing attention whether in studies or practice, where various services MESS can provide are demonstrated. For example, for a normally operating DS, MESS can be scheduled to provide the multi-site reactive power support and reduce the power losses [17]. MESS, mounted on trains as imagined in [18], can also be integrated into the unit commitment problem to boost the economics of bulk grid. During emergency conditions, *e.g.*, if extensive faults occur in a DS, after being fully charged at the site that is still powered, MESS can be further deployed as temporary power source to energize the customers in blackout [19]. This can also be jointly scheduled along with network reconfiguration or/and repair tasks [6], [20]. In addition, through moving among microgrids electrically isolated from each other, MESS can help to coordinate the energy resources in a wider range to achieve a better overall performance (*e.g.*, higher resilience) for the microgrids [7]. In real world, increasing efforts have also been made to promote practical applications of MESS. An increasing number of MESS providers have appeared on the market, such as the Nomad Transportable Power Systems, Alfen, Aggreko, Power Edison, RES [21]-[25]. Customers can obtain MESS solutions whether by renting or purchasing from them. For example, Greener, a clear energy service provider located in Netherlands, ordered forty-three 336 kW·h lithium-ion MESS from Alfen by 2020, and Greener has put them to use in practical projects with various purposes, *e.g.*, to provide backup power for large matches and restaurants, and to solve the issue of voltage dips during live streams [26]. EPRI also launched a research project in 2020 to demonstrate and evaluate the performance of MESS in applications [27]. In addition, heartening signs also appear in terms of relevant policies. For example, in 2018, Massachusetts enacted An Act to Advance Clean Energy (Chapter 227, Acts of 2018). The Act clearly shows interests and puts forward recommendation for studying MESS in the Section 22 of it [28]. Driven by this, in 2020, Massachusetts Department of Energy Resources has made a remarkable and comprehensive research report to recognize the competence of MESS, in which some major practical issues about MESS, including the state of the art in products, specifics for deployment, applications in normal and emergency conditions, and potential advantages, were well addressed [29].

As an improvement of MESS, we have proposed a novel idea, the separable MESS (SMESS) solution, in our recent work [30], whereby the energy storage modules (Mods) and the carriers (Carrs) are scheduled independently, rather than be "fastened" together in the common MESS solution, to obtain extended feasibility and thus better performance. SMESS is a more generalized form of MESS and has greater potential that is worth tapping. In the following part of this paper, we will show how SMESS can work to support the survivability of an IDS. When an IDS is formed, SMESS can be quickly deployed to it from the depot thanks to its mobility. What's more, SMESS will provide an alternative way to rebuild the lifeline for electricity prior to repair of the faulted links, by letting the Mods absorb electricity from the external available source and then be transported by the Carrs into the IDS to release power. It is worth expecting that, by repeating such a process in a proper manner, SMESS can provide a sustainable supply in terms of power and energy to feed the customers and thus support the IDS to survive.

Then, following a sustainable supply provided by SMESS, another concern should be noted, as we previously stated: the available power and energy supplied by SMESS may not be always sufficient to fulfill the original demand in the IDS, due to limited capacity of Mods and limited energy they absorb from external source, like a fluctuant REG. This mismatch possibly puts the IDS under some operating stress regarding power and energy shortages. In this regard, demand response (DR) has proven to be a flexible and useful tool to provide potential reserve and relieve various operating stress by adjusting the demand in allowable range. For example, after faults cause power cut to the downstream loads in a radial DS, DR can be executed in an appropriate manner to curtail some loads to release extra spare capacity of the adjacent healthy feeder, enabling the feeder to further power the area in blackout by closing the tie-line between them [9]. In [11], the collected field data shows that managing the daily power consumption to match the REG generation contributes to keeping the battery in good condition even after days. Ref. [31] integrates DR, as well as electric vehicles and energy storage systems, into the frequency regulation of microgrid to ensure stable operation when the microgrid suffers an unplanning islanding event. In [32], two DR schemes are adopted as corrective actions to eliminate thermal and voltage violations of DS after contingency events, and results show that, compared with the DR scheme whereby all loads that participate in DR are curtailed once the DR is triggered, a smarter DR scheme whereby the loads are optimally curtailed considering the priority and the real-time demand level of DS yields better performance represented by the reduced unnecessary load curtailment. In addition to studies in literatures, DR has also received much attention in practice, *e.g.*, the "Emergency Demand Response Program" and the "Installed Capacity – Special Case Resources Program" administered by the New York Independent System Operator (NYISO), both of which are applied to reduce the system demand during emergency conditions like reserve deficiencies [33], [34].

The failure of the IDS to survive a disaster will undoubtedly cause great losses whether to the IDS itself or to its customers, especially as electricity has become an indispensable resource in more and more fields and activities. Thus, the survival should be a vital task not just of the IDS but also of the customers, requiring efforts by both of them and a strong collaboration with each other. From the above, SMESS and DR, which are expected to provide a sustainable supply and a compensation to the mismatch between supply and demand, respectively, may work well in a coordinated way to boost the survivability of the IDS and this is investigated in this paper. To our best knowledge, a joint scheduling or coordination of DR and SMESS has not been considered yet. Based on the above description, a two-stage robust optimization (RO) model, involving the coordinated scheduling of SMESSs and DR and considering the uncertainty of REG output and loads, is proposed and solved by the column-and-constraint generation (C&CG) method. The main contributions are briefly described as follows:

*1)* SMESS and DR are jointly scheduled to enhance the survivability of the IDS through the coordination of them. Specifically, a two-pronged strategy is developed for this: SMESSs successively transport energy from the external available sources with spare power to the IDS through round trips, while DR is executed inside in a coordinated way to reduce the power demand to relieve the operating stress of the IDS, for example, to reduce the demand to match the available power supply, or to save energy consumption for later use to cope with the future peak demand especially when the available energy imported by SMESSs is to be limited, as will be shown in the case studies. Compared with the cases where only one of SMESS and DR is used, the joint scheduling of SMESS and DR can realize the most demand picked up in the IDS. Especially, even though there is no generator in the IDS for backup use, SMESS and DR can also make the IDS survive with some level of demand supplied continuously. *2)* The energy payback effect following each DR is considered in the scheduling by introducing relevant variables and constraints to represent the energy accumulated during the execution of a DR and the energy to be rebounded when the DR ends. This approach can realize a generalized formulation to model any energy payback effect following a DR as long as the effect can be expressed by the known payback ratios and duration. Finally, the RO model is solved by C&CG method with supports of dual theory and conversion of bilinear terms into linear one.

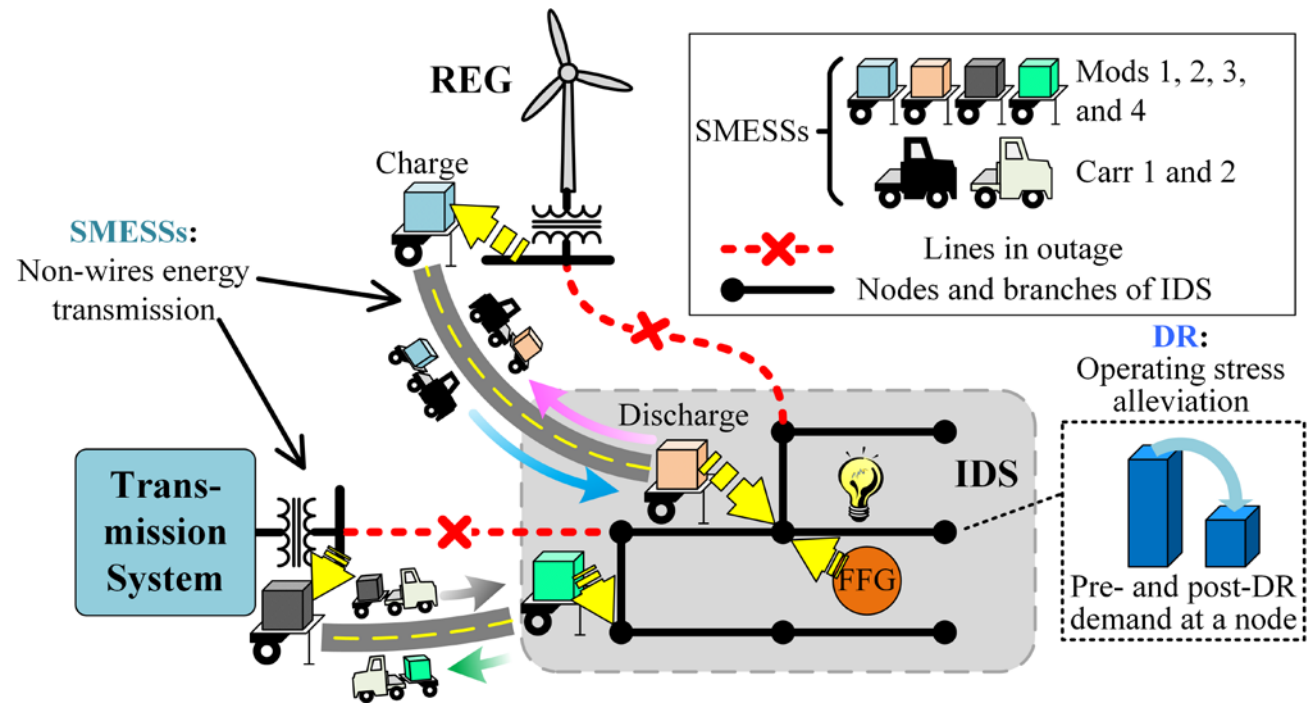

Fig. 1. Illustration of "Keeping the lights on" for an IDS via SMESSs and DR.

The remainder of this paper is organized as follows. Section II provides a brief description of the survivability-oriented strategy; Section III proposes the two-stage RO model; Section IV describes the method to solve the model; Section V provides numerical studies; Section VI provides a discussion about some technical details; and Section VII concludes this paper.

## II. The Survivability-Oriented Strategy

A general scenario of IDS is shown in Fig. 1, where the local area customers lose the continuous supply from the normal power sources (*i.e.*, the substation and the REG) but have back-up small-capacity fossil-fuel-based generation (FFG) within the IDS. In addition, we further assume such an extreme condition that limited fuel is stored in the IDS without any supplement from outside. This scenario can be simply revised to represent any other required scenarios, such as a case where an IDS, which is normally supplied only by the REG in a remote area, loses the supply from the REG by removing the substation node. The model in the following sections can also be simply revised accordingly. Then, a two-pronged strategy to enhance the survivability of the IDS is described as follows:

*1)* From the IDS's external point of view, SMESSs are scheduled to construct non-wires links for energy transmission from the outside "stranded" sources to the IDS. In addition, SMESSs can even realize a continuous power supply for the IDS, provided output of the Mods and traveling behavior of the Carrs are well scheduled.

*2)* From the IDS's internal point of view, DR is scheduled to relieve the energy and power shortages that may arise in the operation of IDS by reducing the demand in the allowable range. Considering that rapid response to the DR request from the IDS operator is beneficial and expected under such an emergency circumstance, in our strategy, the fully dispatchable DR is used, *e.g.*, direct load control (DLC), which can be executed directly by the operator, as in [11] and [32].

Through joint optimization made in the following content, the coordination of the above two approaches can be realized to provide a new way to support the survival of an IDS.

## III. Robust Scheduling Model Formulation

### A. Objective Function

The objective function in (1) is to minimize the following three terms: 1) the total fuel consumed by Carrs for moving and FFGs for generation; 2) the customers' demand reduction due to DR; and 3) the demand not picked up, *i.e.*, the demand of the customers "abandoned" in the scheduling.

$$\min_{\boldsymbol{y}} \kappa_1 \frac{\psi_{11}}{\psi_{1,\max}} + \max_{\boldsymbol{u}\in\mathcal{U}} \min_{\boldsymbol{x}} \kappa_1 \frac{\psi_{12}}{\psi_{1,\max}} + \kappa_2 \frac{\psi_2}{\psi_{2,\max}} + \kappa_3 \frac{\psi_3}{\psi_{3,\max}} \tag{1}$$

where $\boldsymbol{y}$ represents the first-stage decision variables regarding the traveling behaviors of SMESSs and the states of picking up nodes and executing DR and expressed as $\boldsymbol{y}=\{x_{j,i,t}$, $v_{j,i,t}$, $S_{j,t}$, $R_{j,t}$, $\omega_{j,t}$, $\zeta_{k,i,t}$, $\gamma_{k,j,t}$, $\alpha_{j,i,k,t}$, $c_{k,i,t}$, $d_{k,i,t}$, $\delta_i$, $\rho_{i,t}$, $\Psi_{11}\}$; $\boldsymbol{u}=\{\tilde{P}^{\mathrm{OL}}_{i,t}$, $\tilde{P}^{\mathrm{REG}}_{i,t}\}$ represents the uncertain loads and REG outputs; $\mathcal{U}=\{\mathcal{U}_{\mathrm{L}}$, $\mathcal{U}_{\mathrm{REG}}\}$; and the second-stage variable $\boldsymbol{x}$ contains the rest of the variables except those in the uncertainty sets. $\kappa_1$, $\kappa_2$ and $\kappa_3$ can be determined by decision-makers' preference or the analytic hierarchic process (AHP) [6]. $\Psi_{1,\max}$, $\Psi_{2,\max}$, and $\Psi_{3,\max}$ are introduced for normalization and can be estimated as: $\Psi_{1,\max}=\Sigma_{t\in\mathcal{T}}\,(\Sigma_{j\in\mathcal{M}}\mu_j\Delta t+\Sigma_{i\in\{i|\mathcal{F}(i)\neq\Phi\}}\Sigma_{f\in\mathcal{F}(i)}\sigma_f P^{\mathrm{F}}_{f,\max}\Delta t)$, $\Psi_{2,\max}=\Sigma_{i\in\mathcal{N}_{\mathrm{DR}}}w_i\tau^{\mathrm{DR}}_{i,\max}$ $(\max_{t\in\mathcal{T}}\bar{P}^{\mathrm{OL}}_{i,t})T_{i,\mathrm{DR},\max}\Delta t$, and $\Psi_{3,\max}=\Sigma_{i\in\mathcal{N}}w_i\Sigma_{t\in\mathcal{T}}\bar{P}^{\mathrm{OL}}_{i,t}\Delta t$.

The following constraints express the terms in (1):

$$\psi_{11} = \sum\nolimits_{t\in\mathcal{T}} \sum\nolimits_{j\in\mathcal{M}} \mu_j \sum\nolimits_{i\in\mathcal{N}} v_{j,i,t}\Delta t \tag{2a}$$

$$\psi_{12} = \sum\nolimits_{t\in\mathcal{T}} \sum\nolimits_{i\in\{i|\mathcal{F}(i)\neq\Phi\}} \sum\nolimits_{f\in\mathcal{F}(i)} \sigma_f P^{\mathrm{F}}_{f,t}\Delta t \tag{2b}$$

$$\psi_{11} + \psi_{12} \le \hat{\psi}_1 \tag{2c}$$

$$\psi_2 = \sum\nolimits_{t\in\mathcal{T}} \sum\nolimits_{i\in\mathcal{N}_{\mathrm{DR}}} w_i P^{\mathrm{DR}}_{i,t}\Delta t \tag{2d}$$

$$-M\delta_i \le \chi_i - \sum\nolimits_{t\in\mathcal{T}} \tilde{P}^{\mathrm{OL}}_{i,t} \le 0 \;\;,\;\; 0 \le \chi_i \le M\left(1-\delta_i\right), \forall i\in\mathcal{N} \tag{2e}$$

$$\psi_3 = \sum\nolimits_{i\in\mathcal{N}} w_i \chi_i \Delta t \tag{2f}$$

We assume for simplicity that the same type of fuel (*e.g.*, diesel) is consumed by transportation of Carrs and operation of FFGs, as expressed by (2c). Thus, the same weight is adopted for $\Psi_{11}$ and $\Psi_{12}$, both of which means the fuel consumption.

### B. Constraints for SMESSs

The constraints for the scheduling of SMESSs, first proposed in our recent work [30], are used herein, formulated as follows.

$$\sum\nolimits_{i\in\mathcal{N}_{\mathrm{S}}} x_{j,i,t} + \sum\nolimits_{i\in\mathcal{N}_{\mathrm{S}}} v_{j,i,t} = 1 \;\;,\;\; \forall t\in\mathcal{T}\cup\{0\},\, j\in\mathcal{M} \tag{3a}$$

$$\begin{cases} x_{j,i,t} \ge x_{j,i,t-1} + 1.2\left(v_{j,i,t-1} - v_{j,i,t}\right) + 0.4\left(\sum\nolimits_{i\in\mathcal{N}_{\mathrm{S}}} v_{j,i,t-1} - \sum\nolimits_{i\in\mathcal{N}_{\mathrm{S}}} v_{j,i,t}\right) - 0.8 \\ x_{j,i,t} \le x_{j,i,t-1} + \left(v_{j,i,t-1} - v_{j,i,t}\right) - 0.5\left(\sum\nolimits_{i\in\mathcal{N}_{\mathrm{S}}} v_{j,i,t-1} - \sum\nolimits_{i\in\mathcal{N}_{\mathrm{S}}} v_{j,i,t}\right) + 0.7 \end{cases} \tag{3b}$$
$$,\forall t\in\{\mathcal{T}\cup\{0\}\}\setminus\{D\},\, j\in\mathcal{M}, i\in\mathcal{N}_{\mathrm{S}}$$

$$\begin{cases} S_{j,t} \ge x_{j,i,t-1}\sum\nolimits_{i'\in\mathcal{N}_{\mathrm{S}}} T_{j,ii'} + \sum\nolimits_{i'\in\mathcal{N}_{\mathrm{S}}}\left(v_{j,i',t}T_{j,ii'}\right) - \sum\nolimits_{i'\in\mathcal{N}_{\mathrm{S}}} T_{j,ii'} \;,\forall i\in\mathcal{N}_{\mathrm{S}} \\ S_{j,t} \ge 0 \end{cases} \tag{3c}$$
$$,\forall t\in\mathcal{T},\, j\in\mathcal{M}$$

$$R_{j,t} = R_{j,t-1} + S_{j,t} - \sum\nolimits_{i\in\mathcal{N}_{\mathrm{S}}} v_{j,i,t-1} \;,\forall t\in\mathcal{T},\, j\in\mathcal{M} \tag{3d}$$

$$R_{j,t}/M \le \sum\nolimits_{i\in\mathcal{N}_{\mathrm{S}}} v_{j,i,t} \le R_{j,t} \;,\forall\, t\in\mathcal{T}\cup\{0\},\, j\in\mathcal{M} \tag{3e}$$

$$\begin{cases}\omega_{j,t} \ge \sum_{i\in\mathcal{N}_{\mathrm{S}}} v_{j,i,t-1} + \sum_{i\in\mathcal{N}_{\mathrm{S}}} v_{j,i,t} - 2 + \varepsilon \\ -\left(1-\omega_{j,t}\right) \le v_{j,i,t} - v_{j,i,t-1} \le \left(1-\omega_{j,t}\right),\ \forall i \in \mathcal{N}_{\mathrm{S}}\end{cases} \quad (3f)$$
$$,\forall t \in \mathcal{T}, j \in \mathcal{M}$$

$$x_{j,i_j,0} = 1,\ S_{j,0} = 0,\ R_{j,0} = 0,\ \omega_{j,0} = 0,\ \forall j \in \mathcal{M} \quad (3g)$$

$$\sum_{i\in\mathcal{N}_{\mathrm{S}}} \zeta_{k,i,t} + \sum_{j\in\mathcal{M}} \gamma_{k,j,t} = 1,\ \forall t \in \mathcal{T} \cup \{0\}, k \in \mathcal{K} \quad (4a)$$

$$\sum_{k\in\mathcal{K}} W_k \gamma_{k,j,t} \le A_j,\ \forall t \in \mathcal{T}, j \in \mathcal{M} \quad (4b)$$

$$\zeta_{k,i_k,0} = 1,\ \forall k \in \mathcal{K} \quad (4c)$$

$$\gamma_{k,j,t} \le 1 - \sum_{i\in\mathcal{N}_{\mathrm{S}}} x_{j,i,t},\ \forall t \in \mathcal{T}, j \in \mathcal{M}, k \in \mathcal{K} \quad (4d)$$

$$\gamma_{k,j,t} - \zeta_{k,i,t-1} \le x_{j,i,t} + 1 - x_{j,i,t-1},\ \forall t \in \mathcal{T}, j \in \mathcal{M}, i \in \mathcal{N}_{\mathrm{S}}, k \in \mathcal{K} \quad (4e)$$

$$\begin{aligned}&-\left(\sum_{i\in\mathcal{N}_{\mathrm{S}}} x_{j,i,t-1} + \sum_{i\in\mathcal{N}_{\mathrm{S}}} x_{j,i,t}\right) \le \gamma_{k,j,t} - \gamma_{k,j,t-1} \le \\ &\sum_{i\in\mathcal{N}_{\mathrm{S}}} x_{j,i,t-1} + \sum_{i\in\mathcal{N}_{\mathrm{S}}} x_{j,i,t}\ ,\ \forall t \in \mathcal{T}, j \in \mathcal{M}, k \in \mathcal{K}\end{aligned} \quad (4f)$$

$$\begin{cases}\alpha_{j,i,k,t} \le 1 - x_{j,i,t-1};\ \alpha_{j,i,k,t} \le x_{j,i,t}; \\ \alpha_{j,i,k,t} \le \gamma_{k,j,t-1};\ \alpha_{j,i,k,t} \ge -x_{j,i,t-1} + x_{j,i,t} + \gamma_{k,j,t-1} - 1\end{cases} \quad (4g)$$
$$,\forall t \in \mathcal{T}, i \in \mathcal{N}_{\mathrm{S}}, j \in \mathcal{M}, k \in \mathcal{K}$$

$$\zeta_{k,i,t} \ge \sum_{j\in\mathcal{M}} \alpha_{j,i,k,t},\ \forall t \in \mathcal{T}, i \in \mathcal{N}_{\mathrm{S}}, k \in \mathcal{K} \quad (4h)$$

$$\zeta_{k,i,t} - \zeta_{k,i,t-1} \le \sum_{j\in\mathcal{M}} \alpha_{j,i,k,t},\ \forall t \in \mathcal{T}, i \in \mathcal{N}_{\mathrm{S}}, k \in \mathcal{K} \quad (4i)$$

$$c_{k,i,t} + d_{k,i,t} \le \zeta_{k,i,t},\ \forall t \in \mathcal{T}, k \in \mathcal{K}, i \in \mathcal{N}_{\mathrm{S}} \quad (5)$$

$$\begin{aligned}&0 \le P_{k,i,t}^{\mathrm{c.S}} \le c_{k,i,t} P_{k,\max}^{\mathrm{c.S}},\ 0 \le P_{k,i,t}^{\mathrm{d.S}} \le d_{k,i,t} P_{k,\max}^{\mathrm{d.S}}, \\ &-S_{k,\mathrm{Mod}}\zeta_{k,i,t} \le Q_{k,i,t}^{\mathrm{S}} \le S_{k,\mathrm{Mod}}\zeta_{k,i,t},\ \forall t \in \mathcal{T}, k \in \mathcal{K}, i \in \mathcal{N}_{\mathrm{S}}\end{aligned} \quad (6a)$$

$$\left[\sum_{i\in\mathcal{N}_{\mathrm{S}}} \left(P_{k,i,t}^{\mathrm{d.S}} - P_{k,i,t}^{\mathrm{c.S}}\right)\right]^2 + \left(\sum_{i\in\mathcal{N}_{\mathrm{S}}} Q_{k,i,t}^{\mathrm{S}}\right)^2 \le S_{k,\mathrm{Mod}}^2, \forall t \in \mathcal{T}, k \in \mathcal{K} \quad (6b)$$

$$\begin{aligned}&SOC_{k,t} = SOC_{k,t-1} + \left(e_k^{\mathrm{c}} \sum_{i\in\mathcal{N}_{\mathrm{S}}} P_{k,i,t}^{\mathrm{c.S}} - \sum_{i\in\mathcal{N}_{\mathrm{S}}} P_{k,i,t}^{\mathrm{d.S}} \middle/ e_k^{\mathrm{d}}\right) \Delta t \middle/ E_k\ , \\ &SOC_{k,\min} \le SOC_{k,t} \le SOC_{k,\max}\ ,\quad \forall t \in \mathcal{T}, k \in \mathcal{K}\end{aligned} \quad (6c)$$

$$SOC_{k,0} = soc_{k,0}\ ,\ \forall k \in \mathcal{K} \quad (6d)$$

Constraints (3a) to (3g) describe the travel behaviors of SMESS, or more specifically, the Carrs of SMESS. Constraint (3a) means a Carr can only stay at or travel to a node during each time span, *i.e.*, there is always one and only one $x_{j,i,t}$ or $v_{j,i,t}$ that can be activated. Constraint (3b) restricts the state transition between parking state and traveling state of a Carr. Constraint (3c) recognizes the required travel time at the start of a travel. For example, if Carr $j$ stays at node $i$ during time span $t-1$ and travels to node $i'$ during $t$, then $S_{j,t}$ is forced to be $T_{j,ii'}$, a parameter that can be easily predetermined prior to the scheduling. Constraint (3d) represents the residual travel time required to be consumed. $R_{j,t}$ is equal to $S_{j,t}$ during the first time span of a travel, and as the travel goes on, $R_{j,t}$ decreases gradually. Constraint (3e) enforces a Carr to stay traveling until the residual travel time decreases to 0, *i.e.*, until it has been traveling for the necessary travel time. Constraint (3f) maintains the direction during each travel, *e.g.*, if Carr $j$ is traveling during time spans $t-1$ and $t$, then $\omega_{j,t}$ is forced to be 1 and we further have $v_{j,i,t}= v_{j,i,t-1}$. Constraint (3g) represents the initial values of the variables, especially involving the initial positions of Carrs. Detailed derivation of the above constraints can be found in our prior work [35].

Constraints (4a) to (4i) describe the interactive behaviors between the SMESS and the nodes that support its connection. Constraint (4a) represents that a Mod can only stay on a Carr or at a node during each time span. Constraint (4b) imposes the upper limit on carrying capacity of a Carr. Constraint (4c) sets the initial locations of Mods, *i.e.*, Mod $k$ stays at node $i_k$ prior to the scheduling. Constraint (4d) denotes such assumption made in the model: a node always dominates the Mods located at it, *i.e.*, a Carr does not own any Mod when it is staying at a node. Constraint (4e) describes that, when a Carr leaves from a node, it can carry away some of the Mods at the node. Constraint (4f) keeps the states of Mods regarding a Carr when it is traveling, i.e., when Carr $j$ is traveling during time spans $t-1$ and $t$, we have $\gamma_{k,j,t}=\gamma_{k,j,t-1}$ for any Mod $k$, meaning that Carr $j$ cannot lose any Mod on it or obtain a new one. Constraint (4g) defines the auxiliary variable $\alpha_{j,i,k,t}$ determined by $\alpha_{j,i,k,t}=(\neg x_{j,i,t-1})\wedge x_{j,i,t}\wedge\gamma_{k,j,t-1}$. Based on this, constraint (4h) describes that, if there is a Carr that carries Mod $k$ and arrives at node $i$, then node $i$ obtains Mod $k$. Otherwise, we have to keep $\zeta_{k,i,t}\le\zeta_{k,i,t-1}$, meaning that there is not such a Carr and node $i$ cannot obtain a new Mod $k$ if Mod k was not here during the last time span, or, if Mod $k$ was here, node $i$ may lose it which can be carried away by some Carr. This is realized by (4i).

Then, constraints (5) to (6c) restricts the operation of Mods. Constraint (5) denotes that a Mod can be charged or discharged only if it is located at a node. Constraints (6a) and (6b) bounds the active/reactive power output for each Mod. Clearly, constraint (6b) is formulated in a nonlinear form and will be further converted into an approximately linearized form. Constraint (6c) imposes the limits on the state of charge (SOC) for each Mod. Constraint (6d) sets the initial SOC values.

## C. Constraints for DR

The constraints for DR are formulated as follows. In addition, a DR event is commonly followed by a temporary rebound of the load, referred to as the energy payback effect, which can result from that, *e.g.*, the heating or air conditioning equipment tending to use extra energy to remove the heat gained during the reduced service levels [32], [36]. The energy payback effect following the end of each DR event is considered in this paper.

$$\rho_{i,t} \le \delta_i,\ \forall i \in \mathcal{N}_{\mathrm{DR}}, t \in \mathcal{T} \quad (7a)$$

$$\begin{cases}0 \le P_{i,t}^{\mathrm{DR}} \le M\rho_{i,t} \\ P_{i,t}^{\mathrm{DR}} \le \tau_{i,\max}^{\mathrm{DR}} \tilde{P}_{i,t}^{\mathrm{OL}} \\ \tau_{i,\min}^{\mathrm{DR}} \tilde{P}_{i,t}^{\mathrm{OL}} - P_{i,t}^{\mathrm{DR}} \le M\left(1-\rho_{i,t}\right)\end{cases},\ \forall i \in \mathcal{N}_{\mathrm{DR}}, t \in \mathcal{T} \quad (7b)$$

$$\begin{aligned}&\rho_{i,t} = \rho'_{i,D+t},\ \forall i \in \mathcal{N}_{\mathrm{DR}}, t \in \left\{t \middle| h \le t \le 0, t \in \mathcal{Z}\right\}, \\ &\text{where } h = \min\left\{-T_{i,\mathrm{DU},\max}+1, -T_{i,\mathrm{IN},\min}+1\right\}\end{aligned} \quad (7c)$$

$$\begin{aligned}&\sum_{h=0}^{T_{i,\mathrm{DU},\max}} \rho_{i,t+h} \le T_{i,\mathrm{DU},\max},\ \forall i \in \mathcal{N}_{\mathrm{DR}}, \\ &t \in \left\{t \middle| -T_{i,\mathrm{DU},\max}+1 \le t \le D - T_{i,\mathrm{DU},\max}, t \in \mathcal{Z}\right\}\end{aligned} \quad (7d)$$

$$\begin{aligned}&\sum_{h=0}^{T_{i,\mathrm{DU},\min}-1} \rho_{i,t+h} \ge \left(\rho_{i,t} - \rho_{i,t-1}\right) T_{i,\mathrm{DU},\min},\ \forall i \in \mathcal{N}_{\mathrm{DR}}, \\ &t \in \left\{t \middle| -T_{i,\mathrm{DU},\min}+2 \le t \le D - T_{i,\mathrm{DU},\min}+1, t \in \mathcal{Z}\right\}\end{aligned} \quad (7e)$$

$$\begin{aligned}&\sum_{h=0}^{T_{i,\mathrm{IN},\min}-1} \left(1-\rho_{i,t+h}\right) \ge \left(\rho_{i,t-1} - \rho_{i,t}\right) T_{i,\mathrm{IN},\min},\ \forall i \in \mathcal{N}_{\mathrm{DR}}, \\ &t \in \left\{t \middle| -T_{i,\mathrm{IN},\min}+2 \le t \le D - T_{i,\mathrm{IN},\min}+1, t \in \mathcal{Z}\right\}\end{aligned} \quad (7f)$$

$$\sum_{t\in\mathcal{T}} \rho_{i,t} \le T_{i,\mathrm{DR},\max},\ \forall i \in \mathcal{N}_{\mathrm{DR}} \quad (7g)$$

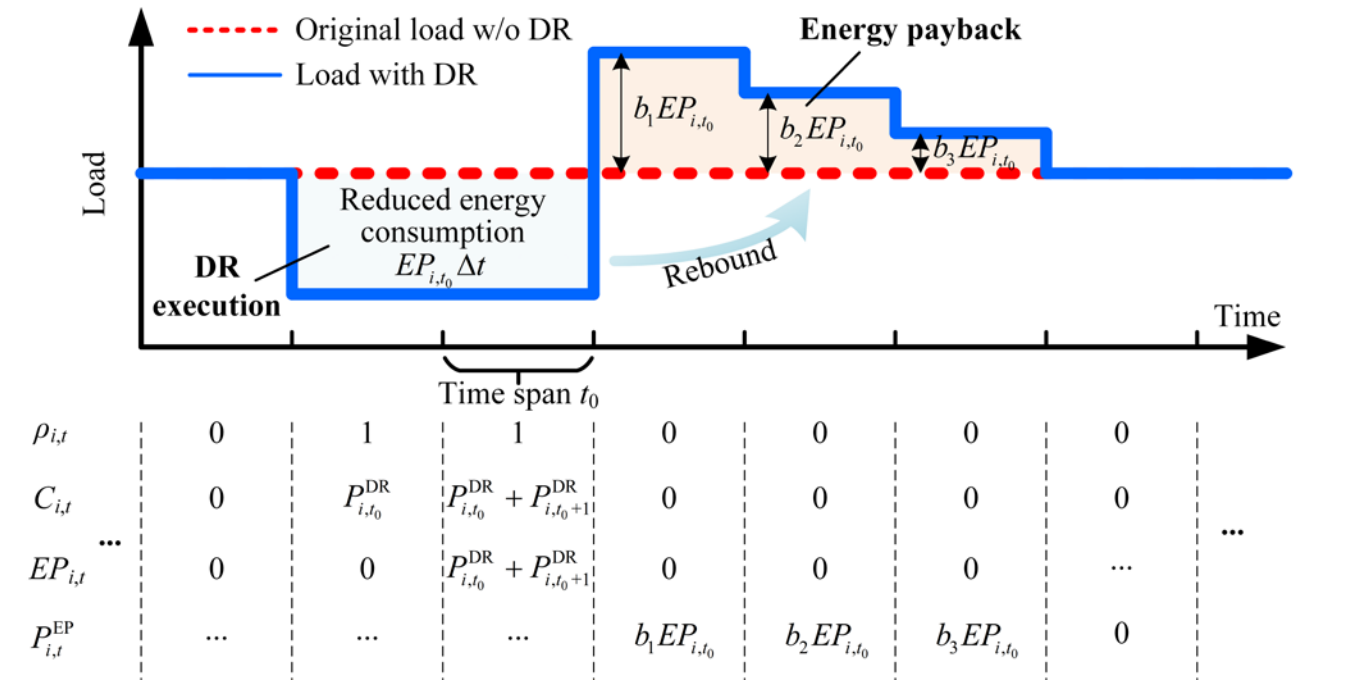

Fig. 2. Illustration about the effect of energy payback and how constraints (8) work. $T_{i,\text{pdu}}$=3Δ$t$ is assumed.

$$\begin{cases} 0 \le C_{i,t} \le M\rho_{i,t} \\ -M\left(1-\rho_{i,t}\right) \le C_{i,t} - \left(C_{i,t-1} + P_{i,t}^{\text{DR}}\right) \le 0\,, \forall i \in \mathcal{N}_{\text{DR}},\ t \in \mathcal{T} \\ C_{i,0} = C'_{i,D} \end{cases} \tag{8a}$$

$$\begin{cases} -M\rho_{i,t+1} \le EP_{i,t} - C_{i,t} \le 0 \\ 0 \le EP_{i,t} \le M\left(1-\rho_{i,t+1}\right) \end{cases} , \forall i \in \mathcal{N}_{\text{DR}}, t \in \left\{\mathcal{T} \setminus \{D\}\right\} \cup \{0\} \tag{8b}$$

$$EP_{i,t} = EP'_{i,D+t}, \forall i \in \mathcal{N}_{\text{DR}}, t \in \left\{t \middle| -T_{i,\text{pdu}} + 1 \le t \le -1\,, t \in \mathcal{Z}\right\} \tag{8c}$$

$$P_{i,t}^{\text{EP}} = \sum_{h=1}^{T_{i,\text{pdu}}} b_{i,h} EP_{i,t-h}\ ,\ \forall i \in \mathcal{N}_{\text{DR}}, t \in \mathcal{T} \tag{8d}$$

Constraints (7a) to (7g) represent the common constraints to DR executions. Specifically, constraint (7a) denotes that where DR can be executed. As indicated by $\delta_i$, we assume that in each scheduling, rather than all the loads that must be supplied, the IDS operator can determine which load is picked up or abandoned. By (7a), DR can be executed at a node only if this node is picked up or being powered. Constraint (7b) imposes upper and lower bounds on the load reduction by DR and also enforces the reduction to be 0 if DR is not executed. Constraint (7c) conducts a link between the decisions over the current and the previous scheduling horizons. Such a link is necessary because, for example, if a DR was still being executed at the end of the previous scheduling, then this must be considered in the current scheduling to ensure that the total duration of this DR, which will cross the two adjacent scheduling horizons, is within the allowable range. Constraints (7d) and (7e) impose upper and lower limits on the duration of a single DR event, respectively. Constraint (7f) imposes a lower limit on the interval between two adjacent DR executed at the same load, *i.e.*, after executing a DR at a load, the operator has to wait at least a certain amount of time to execute the next DR at this load. Constraint (7g) imposes an upper limit on the total duration of all DR events executed at a load during the scheduling horizon.

Then, constraints (8a) to (8d) express the energy payback effect that follows each DR event, and the functions of them are depicted in Fig. 2. The idea of constructing these constraints is described as follows.

Specifically, the variable "$C_{i,t}$" accumulates the reduced energy consumption during a single DR event. In addition, the length of a time span is a constant value, and thus we ignore it when we call this "energy". As realized by (8a), for node $i$, if no DR is being executed during time span $t$, we have $C_{i,t}$=0; otherwise, if a DR is being executed, then $C_{i,t}$ is equal to $C_{i,t}$ + $P_{i,t}^{\text{DR}}$ and thus we realize the accumulation.

Then, the variable "$EP_{i,t}$" identifies the total accumulated one at the end of the DR, part or all of which is about to rebound back into the customer's post-DR demand. We construct (8b) to realize this. It can be easily concluded from (8b) that, only during the final time span of a DR event, $EP_{i,t}$ is equal to $C_{i,t}$, which represents the total reduced energy consumption due to this DR. For other conditions, $EP_{i,t}$ is always equal to 0. Then, constraint (8d) finally realizes the payback, *i.e.*, the total reduced energy consumption resulted from a DR event is rebounded in the subsequent $T_{i,\text{pdu}}$ time spans, at the ratio of $b_{i,1}$, $b_{i,2}$, …. Typically, a 100% payback can be applied to residential customers and 50% to commercial and industrial customers [32], which implies that $\Sigma_{h=1}^{T_{i,\text{pdu}}} b_{i,h}$ is equal to 1 and 0.5, respectively.

In addition, constraint (8c), as well as the third term of (8a), builds a link between the decisions over the current scheduling horizon and the previous one, as does (7c).

## D. *Constraints for FFGs*

The constraints for FFGs operation are given as follows:

$$0 \le P_{f,t}^{\text{F}} \le P_{f,\max}^{\text{F}}\,, 0 \le Q_{f,t}^{\text{F}} \le Q_{f,\max}^{\text{F}}\,, \left(P_{f,t}^{\text{F}}\right)^2 + \left(Q_{f,t}^{\text{F}}\right)^2 \le S_{f,\text{FFG}}^2 \\ , \forall f \in \bigcup_{i\in\mathcal{N}} \mathcal{F}(i), t \in \mathcal{T} \tag{9}$$

Constraint (9) bounds the power output of each FFG. The nonlinear term in (9) will be converted into the linear form approximately, as we will do for (6b).

## E. *Constraints for IDS*

The constraints for IDS operation are formulated as follows based on the linearized DistFlow model [6], [9], [37].

$$\begin{cases} -M\left(1-\delta_i\right) \le P_{i,t}^{\text{L}} - \left(\tilde{P}_{i,t}^{\text{OL}} - P_{i,t}^{\text{DR}} + P_{i,t}^{\text{EP}}\right) \le 0\ , \\ \forall i \in \mathcal{N}_{\text{DR}} \\ -M\left(1-\delta_i\right) \le P_{i,t}^{\text{L}} - \tilde{P}_{i,t}^{\text{OL}} \le 0\ , \forall i \in \mathcal{N} \setminus \mathcal{N}_{\text{DR}} \\ 0 \le P_{i,t}^{\text{L}} \le M\delta_i\ ,\ \forall i \in \mathcal{N} \end{cases} , \forall t \in \mathcal{T} \tag{10a}$$

$$Q_{i,t}^{\text{L}} = \eta_i P_{i,t}^{\text{L}}\ ,\ \forall i \in \mathcal{N}, t \in \mathcal{T} \tag{10b}$$

$$P_{i,t}^{\text{IN.S}} = \sum_{k\in\mathcal{K}} \left(P_{k,i,t}^{\text{d.S}} - P_{k,i,t}^{\text{c.S}}\right),\ Q_{i,t}^{\text{IN.S}} = \sum_{k\in\mathcal{K}} Q_{k,i,t}^{\text{S}}\,, \forall i \in \mathcal{N}_{\text{S}}, t \in \mathcal{T} \tag{11a}$$

$$P_{i,t}^{\text{IN.S}} = 0\,, Q_{i,t}^{\text{IN.S}} = 0\,, \forall i \in \left\{\mathcal{N} \cup \mathcal{S}\right\} \setminus \mathcal{N}_{\text{S}}, t \in \mathcal{T} \tag{11b}$$

$$-P_{i,t}^{\text{IN.S}} \le P_{i,\max}^{\text{sub}}\ ,\ \forall i \in \mathcal{S}_{\text{S}}, t \in \mathcal{T} \tag{11c}$$

$$-P_{i,t}^{\text{IN.S}} \le \tilde{P}_{i,t}^{\text{REG}}\ ,\ \forall i \in \mathcal{S}_{\text{R}}, t \in \mathcal{T} \tag{11d}$$

$$P_{i,t}^{\text{IN.F}} = \sum_{f\in\mathcal{F}(i)} P_{f,t}^{\text{F}}\ ,\ Q_{i,t}^{\text{IN.F}} = \sum_{f\in\mathcal{F}(i)} Q_{f,t}^{\text{F}} \\ , \forall i \in \left\{i \middle| \mathcal{F}(i) \ne \Phi\right\}, t \in \mathcal{T} \tag{12a}$$

$$P_{i,t}^{\text{IN.F}} = 0\ ,\ Q_{i,t}^{\text{IN.F}} = 0\,, \forall i \in \left\{i \middle| \mathcal{F}(i) = \Phi\right\}, t \in \mathcal{T} \tag{12b}$$

$$\sum_{(i',i)\in\mathcal{L}} P_{i'i,t} + P_{i,t}^{\text{IN.S}} + P_{i,t}^{\text{IN.F}} - P_{i,t}^{\text{L}} = \sum_{(i,i')\in\mathcal{L}} P_{ii',t}\ , \\ \sum_{(i',i)\in\mathcal{L}} Q_{i'i,t} + Q_{i,t}^{\text{IN.S}} + Q_{i,t}^{\text{IN.F}} - Q_{i,t}^{\text{L}} = \sum_{(i,i')\in\mathcal{L}} Q_{ii',t}\,, \forall i \in \mathcal{N}, t \in \mathcal{T} \tag{13}$$

$$V_{i',t}^2 = V_{i,t}^2 - 2\left(P_{ii',t} r_{ii'} + Q_{ii',t} x_{ii'}\right)\ , \forall\left(i\,,i'\right) \in \mathcal{L}\,, t \in \mathcal{T} \tag{14}$$

$$V_{i,\min}^2 \le V_{i,t}^2 \le V_{i,\max}^2\,, \forall i \in \mathcal{N}, t \in \mathcal{T} \tag{15}$$

$$P_{ii',t}^2 + Q_{ii',t}^2 \le S_{ii',\max}^2\,, \forall\left(i,i'\right) \in \mathcal{L}\,, t \in \mathcal{T} \tag{16}$$

Constraint (10a) expresses the actual active power load in the IDS considering the effect of DR. By (10a), if node $i$, supposing that it participates in the DR program, is picked up during time span $t$, then its actual load $P^{\mathrm{L}}_{i,t}$ is equal to $\tilde{P}^{\mathrm{OL}}_{i,t} - P^{\mathrm{DR}}_{i,t} + P^{\mathrm{EP}}_{i,t}$; if node $i$ is not a participant of DR and picked up, its actual load is equal to the original value $\tilde{P}^{\mathrm{OL}}_{i,t}$; otherwise, if node $i$ is not picked up, its actual load is forced to be 0. Constraint (10b) denotes the actual reactive power load in the IDS, whereby, for convenience, we impose a fixed power factor on all of the loads. Constraints (11a) represents the power input by SMESS to the nodes in the IDS, and for the nodes that cannot support the connection of SMESS, the power input to them is always 0, as expressed by (11b). Constraints (11c) and (11d) limit the power absorbed by SMESS from external sources, which can be a substation feeder with spare power capacity, or a "stranded" REG caused by the isolation. The external REG may be shut down by the operator when there is no Mod located there and started up when any Mod arrives and required to be charged. Shutting down or starting up the REG can even be done in minutes [38], which is much smaller than the length of a time span, and thus the effect of this time on the scheduling can be ignored. Specific control methods for the REG are out of the scope of this paper and not further discussed here. In addition, the applicability of the proposed strategy is not subject to available REG at all. Clearly, the strategy can work as long as there is spare power capacity outside the IDS for SMESSs to get charged. Constraint (12a) denotes the power input by FFGs, and for nodes without FFG, it is forced to 0 by (12b). Constraint (13) ensures the power balance at each node of the IDS. Constraint (14) represents the relationship of voltage magnitude between two adjacent nodes, and (15) further bounds the voltage magnitude at each node. Finally, constraint (5) restricts the power flows on branches of the IDS.

### *F. Uncertainty Sets*

The uncertainty sets of the REGs' power outputs and the IDSs' loads are given as (17), where the budgets of uncertainty, $\Gamma^{\mathrm{L}}_{i}$ and $\Gamma^{\mathrm{REG}}_{i}$, provide a way to adjust the conservatism of the solution [39]. When all the budgets are equal to 0, a deterministic model without considering any uncertainty is obtained; as the budgets increase, the uncertainty set is enlarged, and the resultant solution is thus increasingly conservative. We follow [40] and assume the budgets as integer.

$$\mathcal{U}_{\mathrm{L}} = \left\{ \tilde{P}^{\mathrm{OL}}_{i,t} \left| \begin{array}{l} \tilde{P}^{\mathrm{OL}}_{i,t} = \overline{P}^{\mathrm{OL}}_{i,t} + \widehat{u}^{\mathrm{OL}}_{i,t} \Delta\widehat{P}^{\mathrm{OL}}_{i.,t} - \breve{u}^{\mathrm{OL}}_{i,t} \Delta\breve{P}^{\mathrm{OL}}_{i.,t} \,, \\ 0 \le \widehat{u}^{\mathrm{OL}}_{i,t} \le 1 \,, 0 \le \widehat{u}^{\mathrm{OL}}_{i,t} \le 1 \,, \forall i \in \mathcal{N}, t \in \mathcal{T} \\ \sum_{t\in\mathcal{T}} \left( \widehat{u}^{\mathrm{OL}}_{i,t} + \widehat{u}^{\mathrm{OL}}_{i,t} \right) \le \Gamma^{\mathrm{L}}_{i} \,, \forall i \in \mathcal{N} \end{array} \right. \right\} \tag{17a}$$

$$\mathcal{U}_{\mathrm{REG}} = \left\{ \tilde{P}^{\mathrm{REG}}_{i,t} \left| \begin{array}{l} \tilde{P}^{\mathrm{REG}}_{i,t} = \overline{P}^{\mathrm{REG}}_{i,t} + \widehat{u}^{\mathrm{REG}}_{i,t} \Delta\widehat{P}^{\mathrm{REG}}_{i.,t} - \breve{u}^{\mathrm{REG}}_{i,t} \Delta\breve{P}^{\mathrm{REG}}_{i.,t} \,, \\ 0 \le \widehat{u}^{\mathrm{REG}}_{i,t} \le 1, 0 \le \breve{u}^{\mathrm{REG}}_{i,t} \le 1, \forall i \in \mathcal{S}_{\mathrm{R}}, t \in \mathcal{T} \\ \sum_{t\in\mathcal{T}} \left( \widehat{u}^{\mathrm{REG}}_{i,t} + \breve{u}^{\mathrm{REG}}_{i,t} \right) \le \Gamma^{\mathrm{REG}}_{i} \,, \forall i \in \mathcal{S}_{\mathrm{R}} \end{array} \right. \right\} \tag{17b}$$

The quadratic terms in (6b), (9) and (16) can be easily converted into linear forms based on the method proposed by [40]. By this method, the circular feasible region of the quadratic constraints is approximately represented by a polygonal region that closely envelops it. Then, the quadratic constraints can be replaced with the linear ones that depict the polygonal region. Specific description about this can be found in [40] or in the electronic appendix of this paper [48]. Thus, all the constraints and the objective function are linear.

**Algorithm 1** C&CG method to solve (17)-(21).

***Step 1***: At first, set $lb(0)=-\infty$ and $ub(0)=+\infty$. Set $\varepsilon$ small enough.

***Step 2***: Solve **MP** and obtain the optimal solution $\{\boldsymbol{y}_k^*, \eta_k^*\}$. Set the lower bound $lb(k)=c^T_{out}\cdot\boldsymbol{y}_k^*+\eta^*$. Specially for $k$=1, we can solve **MP** without considering $\eta$ and (22)-(24).

***Step 3***: Substitute $\boldsymbol{y}_k^*$ into **SP2** and solve it after handling with the bilinear terms as described above. Obtain the optimal solution $\{\boldsymbol{x}_k^*, \boldsymbol{u}_k^*\}$. Set the upper bound $ub(k)=\min\{ub(k-1), c^T_{out}\cdot\boldsymbol{y}_k^*+ c^T_{in}\cdot\boldsymbol{x}_k^*\}$, where $c^T_{in}\cdot\boldsymbol{x}_k^*$ represents the objective value of SP1 or SP2.

***Step 4***: If $ub(k)-\mathrm{lb}(k)<\varepsilon$, then the solving process is completed and return the results. If not, go to ***Step 5***.

***Step 5***: Create variable $\boldsymbol{x}^k$ and add the constraints $\eta\ge c^T_{in}\cdot\boldsymbol{x}^k$, $\boldsymbol{Dy}+\boldsymbol{Ex}^k+\boldsymbol{Fu}_k^*\le\boldsymbol{G}$, and $\boldsymbol{D}_{\mathrm{eq}}\boldsymbol{y}+\boldsymbol{E}_{\mathrm{eq}}\boldsymbol{x}^k=\boldsymbol{G}_{\mathrm{eq}}$ to **MP**. Then, $k=k+1$ and go to ***Step 2***.

## IV. Solution Methodology

The two-stage RO model (1)-(17) can be expressed as the following more compact form and can be solved by the C&CG method [41].

$$\min_{\boldsymbol{y}} \boldsymbol{c}^T_{\mathrm{out}}\boldsymbol{y} + \max_{\boldsymbol{u}\in\mathcal{U}} \min_{\boldsymbol{x}} \boldsymbol{c}^T_{\mathrm{in}}\boldsymbol{x} \tag{17}$$

s.t.

$$\boldsymbol{Ay} \le \boldsymbol{B} \tag{18}$$

$$\boldsymbol{A}_{\mathrm{eq}}\boldsymbol{y} = \boldsymbol{B}_{\mathrm{eq}} \tag{19}$$

$$\boldsymbol{Dy} + \boldsymbol{Ex} + \boldsymbol{Fu} \le \boldsymbol{G} \tag{20}$$

$$\boldsymbol{D}_{\mathrm{eq}}\boldsymbol{y} + \boldsymbol{E}_{\mathrm{eq}}\boldsymbol{x} = \boldsymbol{G}_{\mathrm{eq}} \tag{21}$$

Based on the C&CG method, the model can be solved by iteratively solving the updated master problem and subproblem. Specifically, the master problem in the $k$th iteration is expressed as follows:

**MP:**

$$\min_{\boldsymbol{y}} \boldsymbol{c}^T_{\mathrm{out}}\boldsymbol{y} + \eta$$

s.t. (18), (19)

$$\eta \ge \boldsymbol{c}^T_{\mathrm{in}}\boldsymbol{x}^l \,,\ l = 1, 2, \cdots, k-1 \tag{22}$$

$$\boldsymbol{Dy} + \boldsymbol{Ex}^l + \boldsymbol{Fu}^*_l \le \boldsymbol{G} \,,\ l = 1, 2, \cdots, k-1 \tag{23}$$

$$\boldsymbol{D}_{\mathrm{eq}}\boldsymbol{y} + \boldsymbol{E}_{\mathrm{eq}}\boldsymbol{x}^l = \boldsymbol{G}_{\mathrm{eq}} \,,\ l = 1, 2, \cdots, k-1 \tag{24}$$

where $\boldsymbol{u}^*_l$ is the optimal scenario (*i.e.*, $\boldsymbol{u}^*_l$ represents the worst case) obtained by solving the subproblem in the $l$th iteration.

After obtaining the optimal $\boldsymbol{y}^*_k$ by solving the above **MP**, the subproblem can be written as:

**SP1:**

$$\max_{\boldsymbol{u}\in\mathcal{U}} \min_{\boldsymbol{x}} \boldsymbol{c}^T_{\mathrm{in}}\boldsymbol{x} \tag{25}$$

s.t.

$$\boldsymbol{Dy}^*_k + \boldsymbol{Ex} + \boldsymbol{Fu} \le \boldsymbol{G} \tag{26}$$

$$\boldsymbol{D}_{\mathrm{eq}}\boldsymbol{y}^*_k + \boldsymbol{E}_{\mathrm{eq}}\boldsymbol{x} = \boldsymbol{G}_{\mathrm{eq}} \tag{27}$$

To solve **SP1**, we can equivalently convert the inner linear minimization problem to its dual form based on the strong duality theorem, and then we rewrite **SP1** as

**SP2:**

$$\max_{\boldsymbol{\lambda}_1, \boldsymbol{\lambda}_2, \boldsymbol{u}} \boldsymbol{\lambda}_1^T \left( \boldsymbol{G} - \boldsymbol{Dy}^*_k - \boldsymbol{Fu} \right) + \boldsymbol{\lambda}_2^T \left( \boldsymbol{G}_{\mathrm{eq}} - \boldsymbol{D}_{\mathrm{eq}}\boldsymbol{y}^*_k \right) \tag{28}$$

s.t.

$$\boldsymbol{\lambda}_1^T \boldsymbol{E} + \boldsymbol{\lambda}_2^T \boldsymbol{E}_{\mathrm{eq}} = \boldsymbol{c}^T_{\mathrm{in}} \tag{29}$$

$$\boldsymbol{\lambda}_1 \le 0 \tag{30}$$

$$\boldsymbol{u} \in \mathcal{U} \tag{31}$$

where $\boldsymbol{\lambda}_1$ and $\boldsymbol{\lambda}_2$ are dual variables of the inner problem of **SP1**.

Note that the bilinear term $\boldsymbol{\lambda}_1^T\cdot\boldsymbol{u}$, more specifically, the terms $\lambda_1(n)\cdot\widehat{u}^{\mathrm{L}}_{i,t}$, $\lambda_1(n)\cdot\breve{u}^{\mathrm{L}}_{i,t}$, … where $\lambda_1(n)$ is the $n$th element of $\boldsymbol{\lambda}_1$ if we substitute (17) into (28), makes **SP2** still hard to solve. However, for bilinear programming **SP2**, there exists an optimal solution lying at a vertex of its feasible region [42]. Thus, we can set the budgets $\Gamma_i^{\mathrm{L}}$ and $\Gamma_i^{\mathrm{REG}}$ in (17) as integers and then the optimal $\hat{u}_{i,t}^{\mathrm{L}}$, $\check{u}_{i,t}^{\mathrm{L}}$, $\hat{u}_{i,t}^{\mathrm{REG}}$, and $\check{u}_{i,t}^{\mathrm{REG}}$ belong to $\{0, 1\}$, as proved in [43]. From this, we define $\hat{u}_{i,t}^{\mathrm{L}}$, $\check{u}_{i,t}^{\mathrm{L}}$, $\hat{u}_{i,t}^{\mathrm{REG}}$, and $\check{u}_{i,t}^{\mathrm{REG}}$ as binary variables and the bilinear terms in (28) can be converted to linear forms by introducing new variables and adding new constraints to **SP2**, as in [40]. For example, for $\lambda_1(n)\cdot\hat{u}_{i,t}^{\mathrm{L}}$, we can introduce a new variable $\hat{z}_{n,i,t}^{\mathrm{L}}$ to replace $\lambda_1(n)\cdot\hat{u}_{i,t}^{\mathrm{L}}$ in (28) and add the following constraints to **SP2**:

$$\begin{aligned} &-M\left(1-\hat{u}_{i,t}^{\mathrm{L}}\right) \le \hat{z}_{n,i,t}^{\mathrm{L}} - \lambda_1(n) \le M\left(1-\hat{u}_{i,t}^{\mathrm{L}}\right), \\ &-M\hat{u}_{i,t}^{\mathrm{L}} \le \hat{z}_{n,i,t}^{\mathrm{L}} \le -M\hat{u}_{i,t}^{\mathrm{L}} \end{aligned} \tag{32}$$

Finally, both **MP** and **SP2** are mixed-integer linear programmings (MILPs) and can be solved by off-the-shelf solvers. The specific C&CG method is given as **Algorithm 1**.

## V. Numerical Results

In this section, we conduct case studies to verify the effectiveness of the proposed model. The modified IEEE 33-feeder system is used as the IDS [37]. The model is coded on the MATLAB R2020b platform with the YALMIP toolbox [44] and the MILPs are solved by Gurobi v9.1.1 on a computer with an Intel Core i5 8250U CPU and 12 GB RAM.

### A. *Test System and Scenario*

We focus on the cases where an IDS loses connections to the normal power source for a long time in this paper. A wind-based REG with a rated power of 0.8 MW acts as the main source that powered the IDS under normal circumstances and is assumed to be dropped from the IDS due to some major disaster in the test, as shown in Fig. 3. For simplicity, we assume that no other faults exist on branches or nodes inside the IDS and that the topology of the IDS is fixed during scheduling. Thus, tie lies originally in the test system are removed, given that network reconfiguration is out of our scope. A light demand level is assumed for the IDS and the rated load at each node in the IDS has been shrunk to one-fifth of the original value in [37]. The priority weights of loads are randomly assigned from 1 to 5. Types of loads (commercial or residential) are arbitrarily set and eight of them are selected as participants of DR. The load profiles of Los Angeles from [45] are used to depict the IDS load, and the wind power profile from CAISO [46] is used to depict the REG output in the test. The day-ahead forecasted loads and REG output are drawn in Fig. 4 as multipliers of the rated values. Two FFGs for back-up use are assumed in the IDS, each of which has a 200 kW/250 kVA capacity, as given in [47]. The SMESS in the test comprises one Carr (*e.g.*, a tractor) and two 300 kW/750 kW·h Mods, all of which are initially located at node 1. The initial SOC of the two Mods is set as 0.5. The Carr can carry one or both of the Mods simultaneously, and 1 time span is assumed for it to travel between node 1 and the stranded REG. The budgets of uncertainty in (17) are set as 24. $\kappa_1$, $\kappa_2$, and $\kappa_3$ are determined by AHP, based on the assumption in the test that serving as many loads as possible is far more important than saving the consumed fuel and reducing the DR executions. The main parameters are listed in Table I.

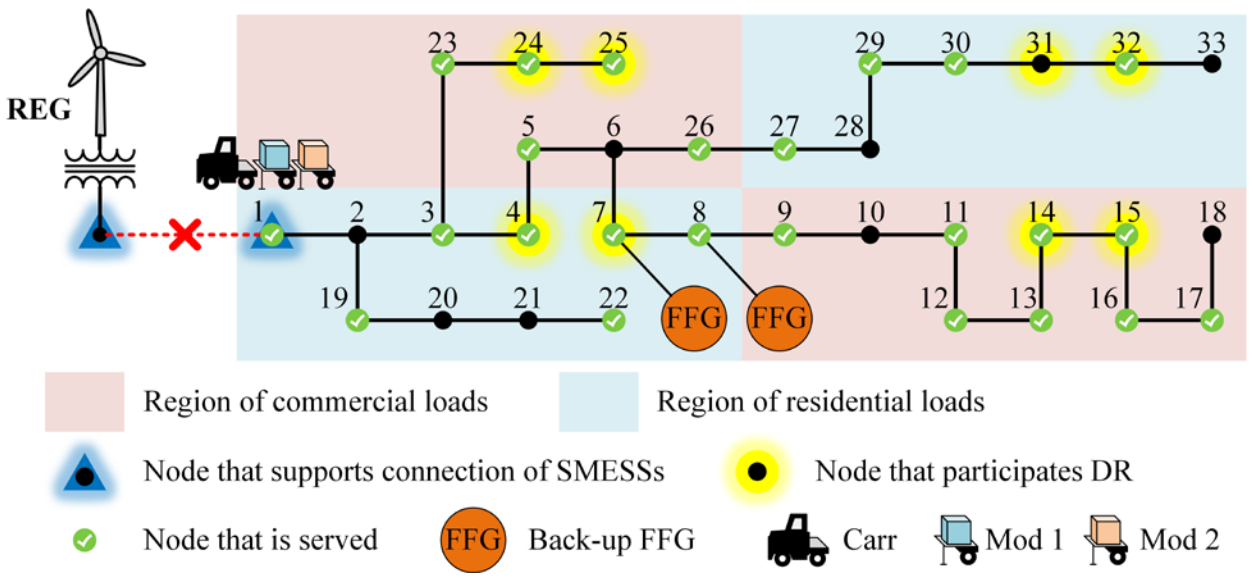

Fig. 3. The test system.

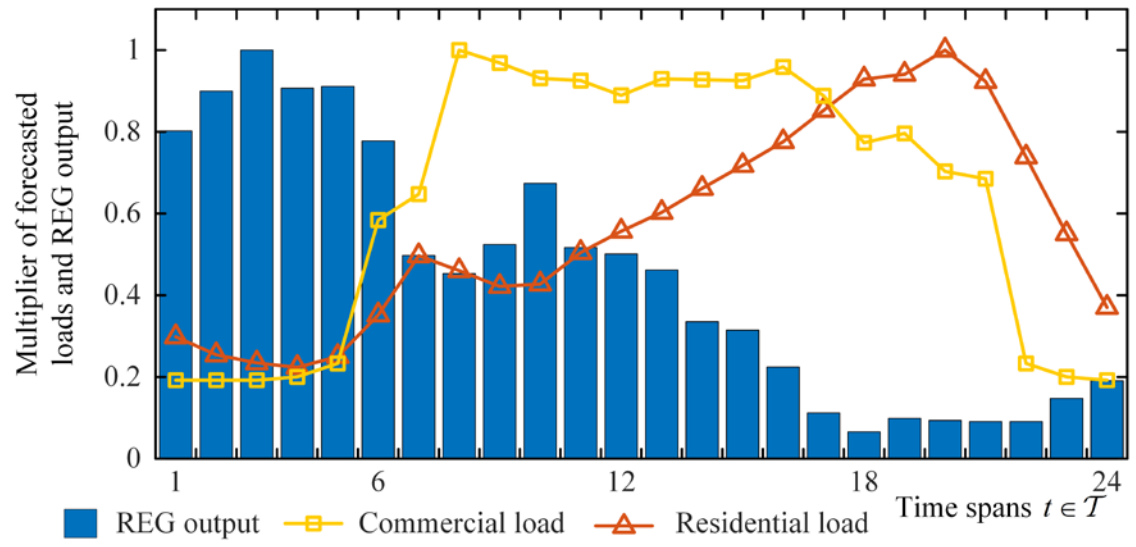

Fig. 4. The REG output and load profile.

TABLE I
SPECIFIC PARAMETERS ADOPTED IN THE TEST

| | | | | |
|---|---|---|---|---|
| **About SMESS** (for $\forall k\in\mathcal{K}$) | $P_{k,\max}^{\mathrm{c.S}}/P_{k,\max}^{\mathrm{d.S}}$ (kW) | 300/300 | $S_{k,\mathrm{Mod}}$ (kVA) | 300 |
| | $e_k^c/e_k^d$ | 0.95/0.95 | $E_k$ (kW·h) | 750 |
| | $SOC_{k,\min}/SOC_{k,\max}$ | 0.1/0.9 | $\mu_j$ $(L/\Delta t)$ | 8 |
| **About DR** (for $\forall i\in\mathcal{N}_{\mathrm{DR}}$) | $T_{i,\mathrm{DU,max}}$ (h) | 4 | $T_{i,\mathrm{DU,min}}$ (h) | 2 |
| | $T_{i,\mathrm{IN,max}}$ (h) | 3 | $T_{i,\mathrm{DR,max}}$ (h) | 8 |
| | $\tau_{i,min}^{DR}/\tau_{i,max}^{DR}$ | 0.4/0.6 | $T_{i,\mathrm{pdu}}$ (h) | 2 |
| | $b_1$, $b_2$ for commercial loads | 0.35, 0.15 | $b_1$, $b_2$ for residential loads | 0.7, 0.3 |
| **About FFGs** (for $\forall f\in\cup_{i\in\mathcal{N}}\mathcal{F}(i)$) | $P_{f,\max}^{\mathrm{F}}$ (kW) | 200 | $Q_{f,\max}^{\mathrm{F}}$ (kVar) | 200 |
| | $S_{f,\mathrm{FFG}}$ (kVA) | 250 | $\sigma_f(L/\mathrm{kW}\cdot\Delta t)$ | 0.282 |
| Uncertainty set | $\Delta\hat{P}_{i,t}^{\mathrm{OL}}$ and $\Delta\check{P}_{i,t}^{\mathrm{OL}}$ | $0.2\bar{P}_{i,t}^{\mathrm{OL}}$ | $\Delta\hat{P}_{i,t}^{\mathrm{REG}}$ and $\Delta\hat{P}_{i,t}^{\mathrm{REG}}$ | $0.2\bar{P}_{i,t}^{\mathrm{REG}}$ |
| **Others** | $\Delta t$ (h) | 1 | $D$ (h) | 24 |
| | $\kappa_1$, $\kappa_2$, $\kappa_3$ | 0.1618, 0.0679, 0.7703 | | |

### B. *Solution and Analysis*

Based on the above parameters, the proposed two-stage RO model is solved after three iterations using the method in Section IV. The obtained first-stage decisions, including the states of nodes being picked up or executed DR and traveling behaviors of SMESSs, are shown in Fig. 3, Fig. 5, and Fig. 7. By substituting the first-stage results and the worst-case scenario obtained from the final iteration into the second-stage problem, *i.e.*, **SP1** while the uncertainty is realized and $u$ is known, the second-stage results under the worst-case scenario, including the power outputs of the Mods and FFGs and the load reduction of DR, are solved and shown in Fig. 5 - Fig. 7.

Twenty-four of the IDS nodes, accounting for approximately 80% of the total demand, are picked up and served during the scheduling, as shown in Fig. 3, while the remaining demand of the other nine nodes is "abandoned". Several round trips of the Mods are completed by the Carr between the IDS and the external REG, as shown in Fig. 5 (a); and as expected, the two Mods are in a charging state when located at the REG and in a discharging state at the IDS to realize the transportation of energy between the two locations. In addition, it is observed from Fig. 5 (b) that, much of the time (during time spans 8 - 21), the two Mods work alternately as the auxiliary source with the FFGs to supply the IDS continuously. For the two FFGs, since

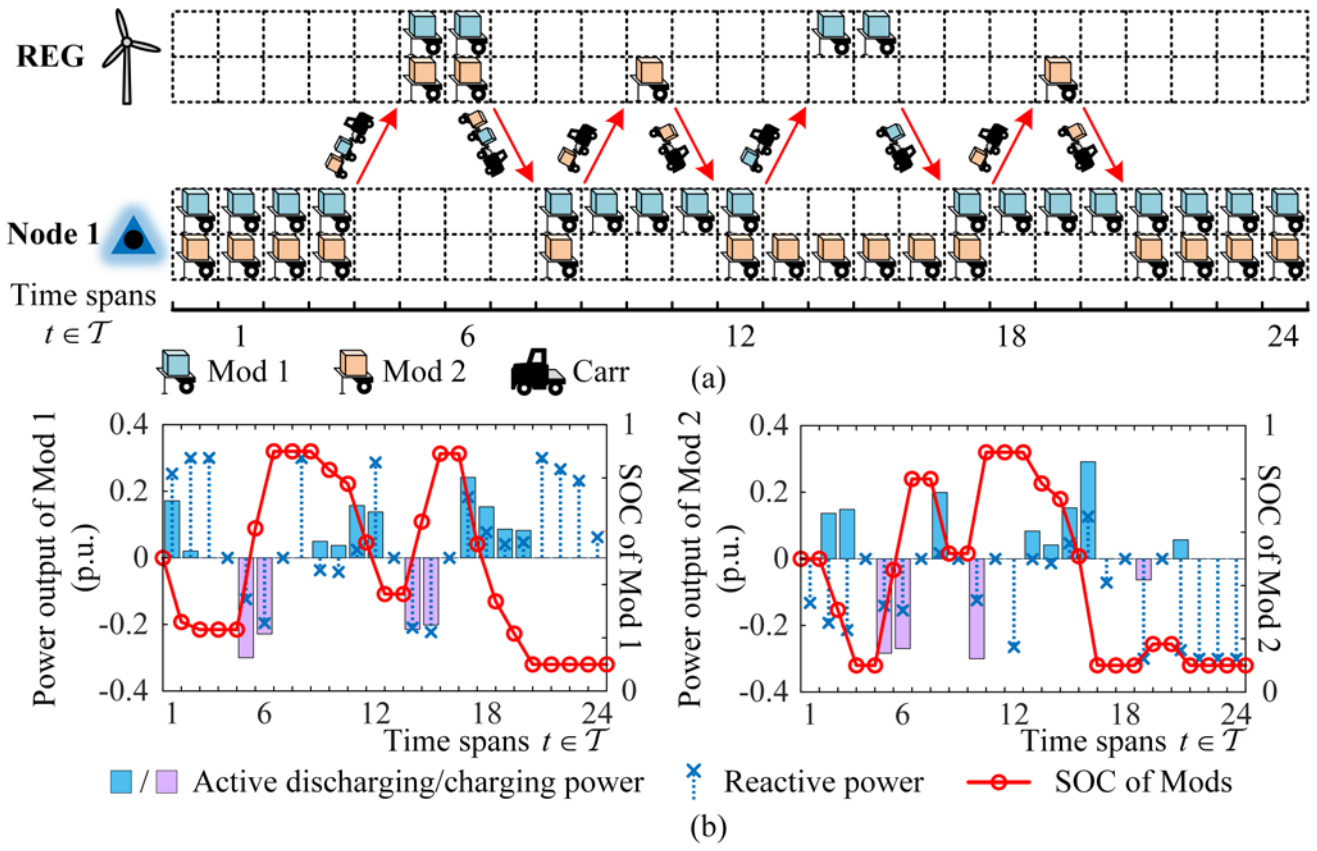

Fig. 5. Scheduling results of (a) traveling behaviors of SMESS; and (b) power outputs and SOC of Mods 1 and 2 under the worst-case scenario. The base power is set as 1 MVA.

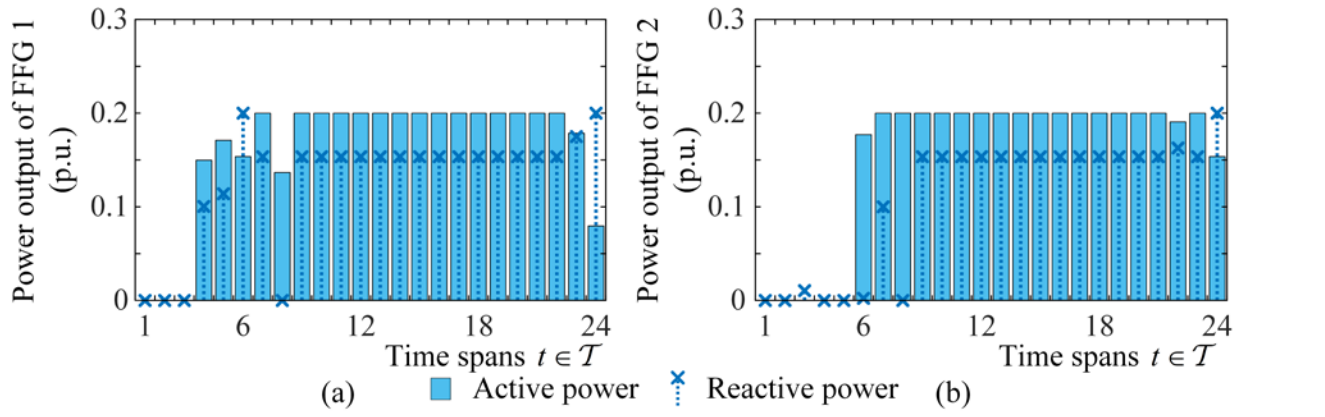

Fig. 6. Scheduling results of power outputs of FFGs at (a) node 7 and (b) node 8 under the worst-case scenario.

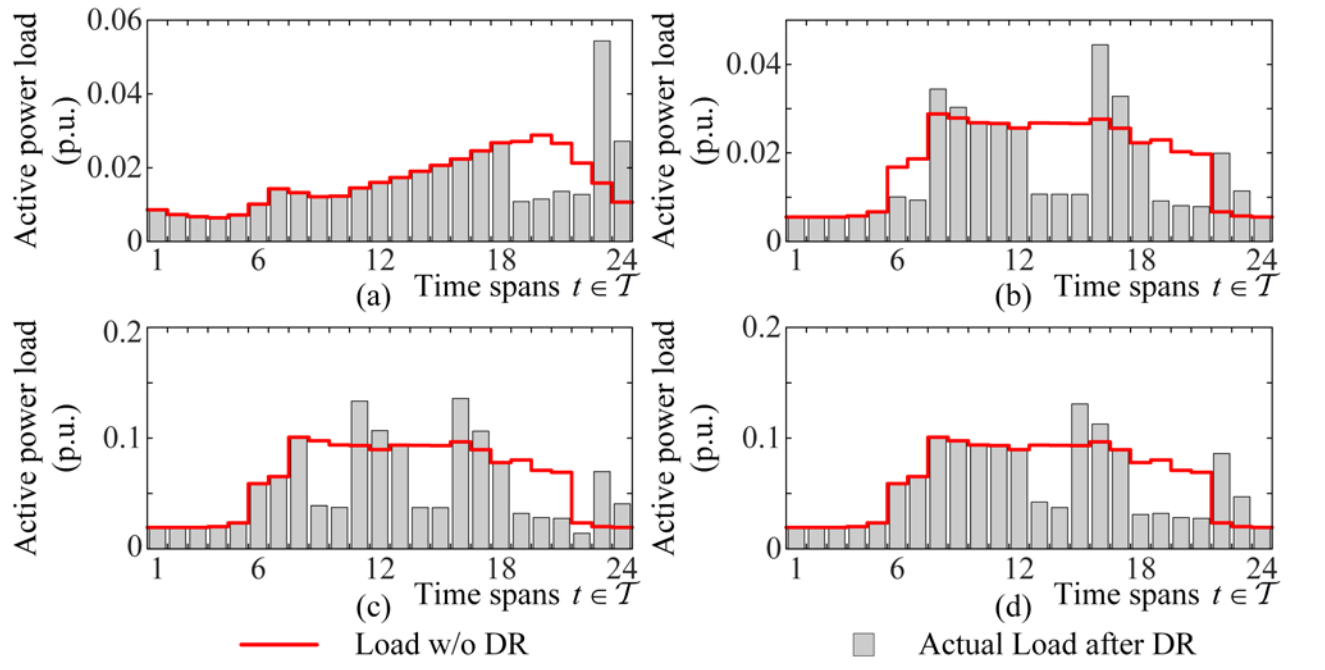

Fig. 7. Scheduling results of DR executions at (a) node 4, (b) node 14, (c) node 24, and (d) node 25.

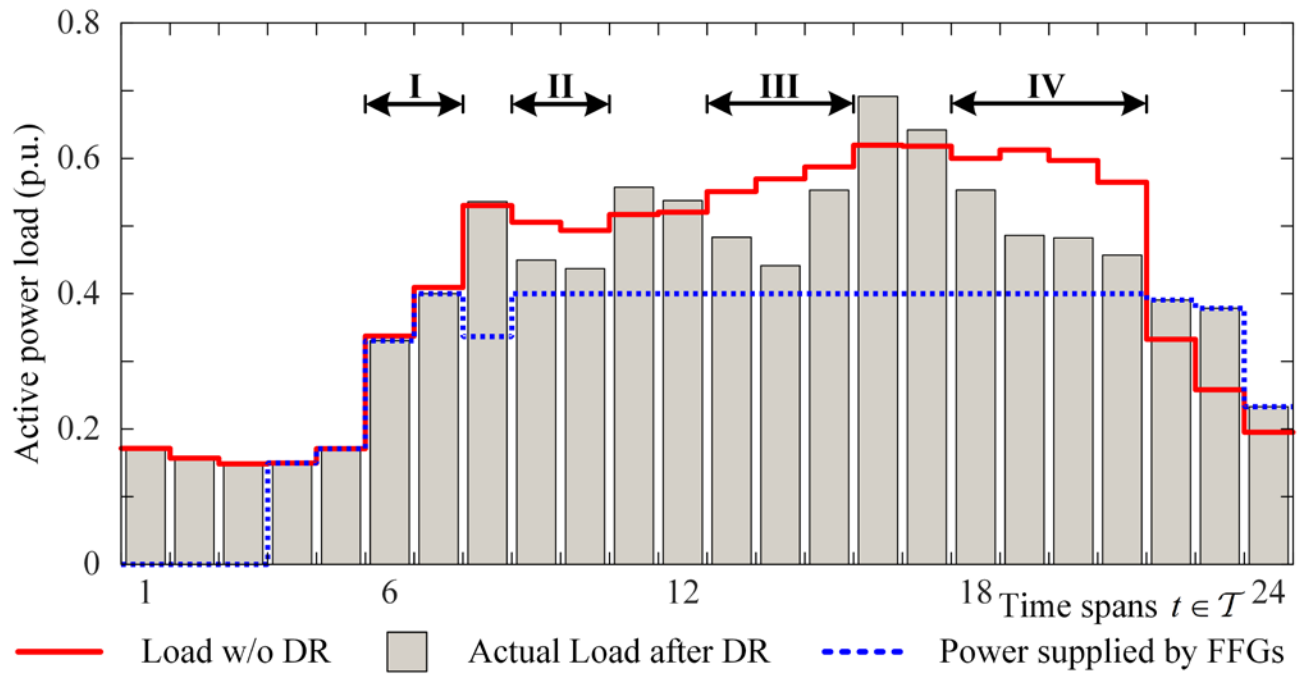

Fig. 8. The total demand served in the IDS without DR and after DR.

the weight value of serving demand $\kappa_3$ is set much higher than that of fuel consumption $\kappa_1$ in the objective function, a full-power output is mostly realized for both of the FFGs, as shown in Fig. 6. DR is executed at nodes 4, 14, 24, and 25. It seems that commercial loads are preferred to the execution of DR due to their lower rebound effect than residential loads. We draw the actual hourly total demand served during the scheduling and its value without DR in Fig. 8, which shows an interesting

TABLE II
COMPARISON OF RESULTS AMONG CASES 1 TO 4

| ***Case*** | $\Psi_{11}/\Psi_{1,\max}$ | $\Psi_{12}/\Psi_{1,\max}$ | $\Psi_2/\Psi_{2,\max}$ | $\Psi_3/\Psi_{3,\max}$ |
|---|---|---|---|---|
| ***1*** | 0.0207 | 0.6924 | 0.4205 | 0.1330 |
| ***2*** | 0.0207 | 0.6184 | 0 | 0.1939 |
| ***3*** | 0 | 0.7813 | 0.1637 | 0.1975 |
| ***4*** | 0 | 0.7455 | 0 | 0.2278 |
| ***Special I*** | 0.0259 | 0.6476 | 0.5136 | 0.1330 |

coordination between SMESS and DR.

From the total load perspective, four periods can be recognized as the load reduction due to DR, as shown in Fig. 8. For *Period I*, a tiny reduction occurs because, if without DR, the power demand would slightly exceed the available power of the two FFGs in time span 7 when the Mods have still been on the trip. Thus, DR is executed at node 14 to cope with that slight power shortage issue. For *Periods II* and *III*, as shown in Fig. 7, DRs are executed by commercial loads, which have a gain of energy payback below 100%, and load reduction occurs mainly resulting from the purpose of saving energy to use for the following peak demand during time spans 16 - 17. Specifically, during *Period II* or time spans 9 - 11, Mod 1 works as the only auxiliary source except for the FFGs. Saving energy is required for Mod 1 because sufficient energy should be kept to confront that peak and supply the IDS after the peak (as shown in Fig. 5, a near full discharge of Mod 1 is observed around time span 18). In addition, even though Mod 1 is carried soon to the REG and charged, during time span 14 - 15, the power output of the REG is limited and below the full charging power of Mod 1 under the obtained worst case where only 80% of the forecasted power is available during this period. During *Period III*, Mod 2 acts as the only auxiliary source. Similarly, saving energy is important for it to confront the upcoming peak demand, and conservative operation is required during this period. Then, after the peak demand, during *Period IV*, the available energy of the two Mods is limited. As shown in Fig. 5 (b), Mod 2 is charged at the REG only to a low level due to the REG's very limited power under the worst case. Both of the Mods use up their energy at the end of this period, and if without DR, as shown by the part between the red line and the blue line in Fig. 8, *Period IV* cannot be successfully rid through due to the greater energy shortage.

## C. *Comparison among Cases*

Based on the above test system, the effectiveness of our proposed method is further demonstrated by comparison among the following cases. *Case 1*: SMESS and DR (*i.e.*, the proposed method and the analysis in the previous subsection). *Case 2*: SMESS without DR. *Case 3*: Stationary Mods and DR. *Case 4*: Stationary Mods without DR. The revisions to the model for realizing the above cases are given in the electronic appendix of this paper [48]. For *Cases 3* and *4*, Mod 1 and Mod 2 are fixed at their initial location, *i.e.*, at node 1. The results of the terms in the objective function under the four cases are given in Table II. By using the proposed method that coordinates the scheduling of SMESS and DR in the IDS operation, the lowest weighted abandoned demand is realized under *Case 1*, which is decreased by 31.4% and 32.7% compared with scheduling SMESS and DR alone under *Case 2* and *Case 3*, respectively, and especially by 41.6% compared with *Case 4*. In brief, comparing *Case 1* to *Case 2* (or *Case 3* to *Case 4*), DR enables more loads to be served by aptly relieving the operating stress in terms of power and energy shortage, as analyzed before, though it also brings more fuel consumed for FFGs generation under the preset $\kappa_1$, $\kappa_2$, $\kappa_3$. By using SMESS, comparing *Case 3* with *Case 1* (or comparing *Case 4* with *Case 2*), more loads are served with less fuel consumption due to the increased available power and the energy supplemented from outside.

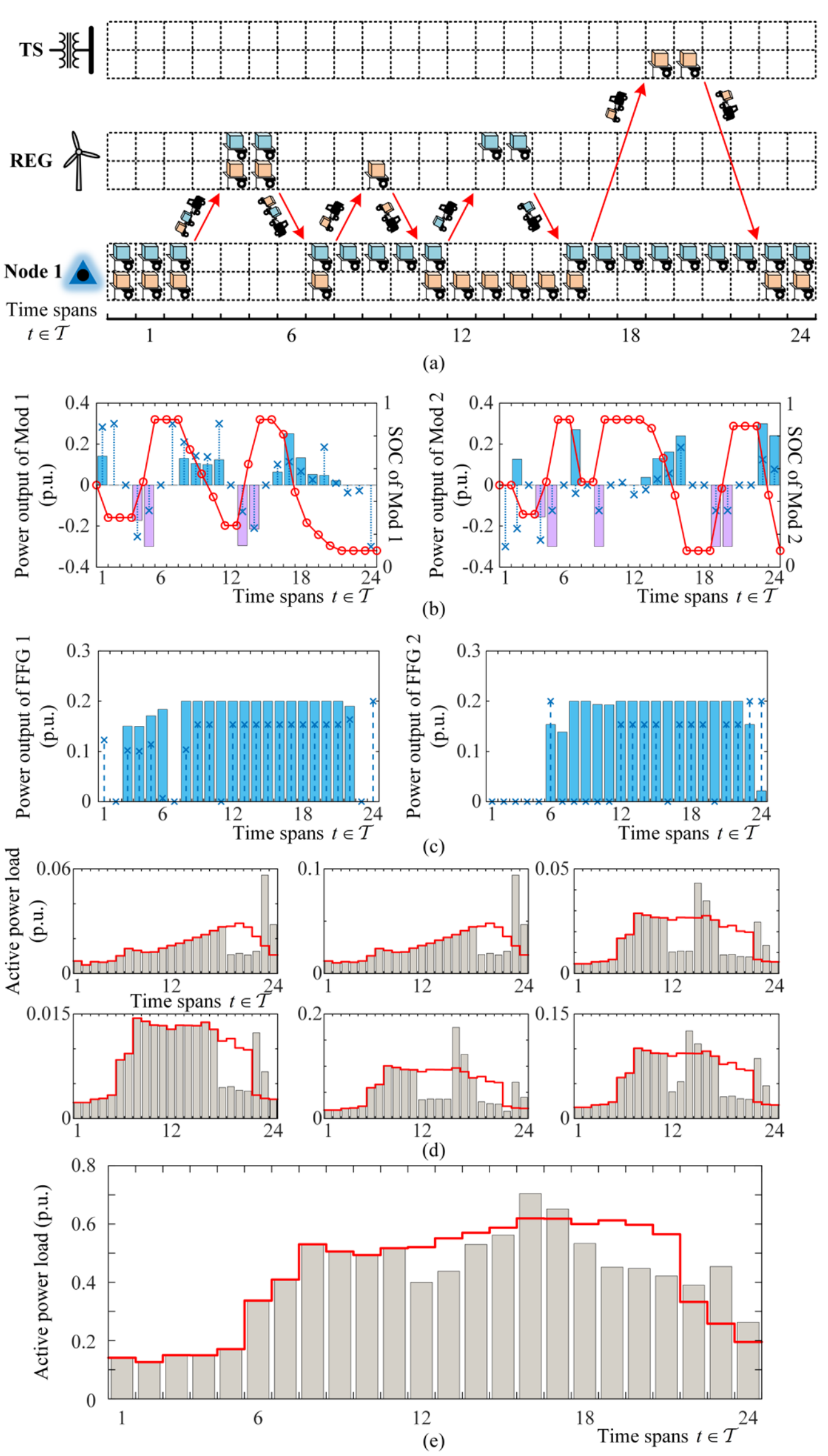

Fig. 9. Scheduling results of (a) traveling behaviors of SMESS, (b) power outputs and SOC of Mods 1 (left) and 2 (right), (c) power outputs of FFGs, (d) DR executions at nodes 4, 7, 14, 15, 24, and 25 (left to right, top to bottom), and (e) the total demand served without DR and after DR under the *Special Case I*. The same legend as Fig. 5 - Fig. 8 is used.

## D. *Special Case I*

An interesting *special case* is then studied: Suppose that, in addition to the REG, there is another available source outside the IDS. The source can be provided by a nearby transmission system, *e.g.*, it may be a health substation node or a DS feeder with adequate spare power. Specifically, the additional source is assumed to have a spare power capacity $P_{i,\max}^{\text{sub}}$ larger the maximum charging power when both of the Mods get charged from it; however, the Carr needs to spend 2 time spans traveling between this source and IDS, which is longer than the travel between the REG and IDS. Then, we solve the model and see what will happen.

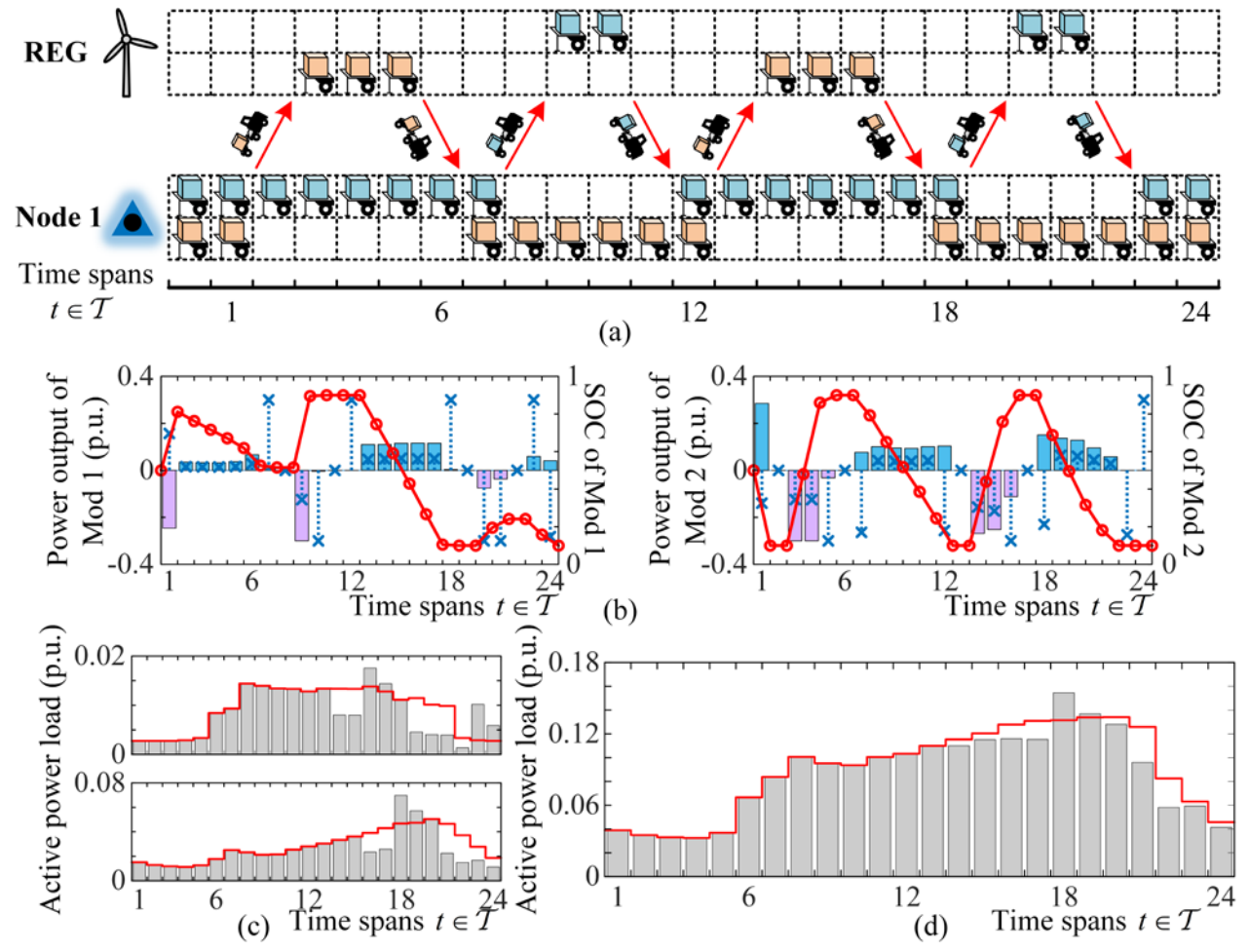

Fig. 10. Scheduling results of (a) traveling behaviors of SMESS, (b) power outputs and SOC of Mods 1 (left) and 2 (right), (c) DR executions at nodes 15 (upper) and 32 (lower), and (d) the total demand served without DR and after DR under the *Special Case II*. The same legend as Fig. 5 - Fig. 8 is used.

The main scheduling results are given in Fig. 9 and the objective value is given in Table II for comparison (see *Special I*). From Fig. 9 (a), we can see the behaviors of SMESS are similar to before during the period prior to the 17th time span. In this period, the REG is chosen for charging the Mods because its power output is still adequate without dropping too much and it is also closer to the IDS. In the condition where adequate power is available outside, a closer position of the external source is clearly important since it enables SMESS to realize a more frequent transportation and thus to input more energy into IDS. However, in the "night" after that, the available power of REG drops down to a so small value that Mod 2 goes to the further source TS to get charged greedily at the full power. Then, Mod 2 is transported back to the IDS with rich energy, which greatly alleviates the demand for the generation of FFGs, as shown in Fig. 9 (c). From the comparison among the results in Table II, the fuel consumption of FFGs is reduced.

However, more DR is executed in this case. As shown by Fig. 9 (d), two other nodes 7 and 15 suffer DR during the scheduling. The DR executed at these two nodes occurs after the 18th time span and contributes to reduce the total power demand, which otherwise would still be at the peak but without sufficient supply because the only Mod in the IDS (Mod 1) is running out of the stored energy as shown in Fig. 9 (b).

## E. *Special Case II*

In addition, another *special case* is given as "SMESS and DR without FFGs", *i.e.*, the IDS losing the FFGs. As shown by the result in Fig. 10 (a) and (b), a continuous power supply to the IDS is realized for the whole time by the alternate work of the two Mods of SMESS. This enables eight of the nodes be served during the scheduling even without FFGs, with the help of DR executed at nodes 15 and 32, as shown in Fig. 10 (c) and (d).

## VI. Discussion

Commercial battery-based MESS solutions have come to the market and the relevant technologies continue to grow. Actually, MESS bears strong resemblance to the common electric vehicle (EV) in terms of the interconnection to power grids.

MESS is desired to provide power support conveniently to the grids at different sites. Such convenience can be well provided by the plug-and-play capability, which enables MESS to immediately start working once on site. Although, in contrast to EV that has been developed for many years and run on the road worldwide, there are much fewer proper industrial standards for the design of MESS and its ancillary facilities, the plug-and-play design is becoming a consensus for MESSs of different providers [21]-[25]. Similar to EVs, most of the state-of-the-art MESSs on the market have an all-in-one structure that has necessary facilities integrated in a container, including battery cells, systems for control, power conversion, heating, ventilation, air-conditioning, circuit breakers, and even a transformer for different voltage-level use [23]. MESS can be easily plugged into the grid through the same physical interface as EV, *i.e.*, the EV service equipment (EVSE) and standard connectors [29]. Besides this, platforms specifically designed for some MESSs also support simple and fast connection or disconnection of them to or from the grids [21]. In addition, the hot-plugging capability recommended to MESSs makes them be connected or disconnected without causing any power interruption of the operating grids [49], [50]. All of the above facilitate the realization of the plug-and-play capability of MESS, by which the connection and disconnection of MESS can be accomplished in seconds or minutes, a much shorter time than the time step in the scheduling, *e.g.*, 1 hour in Section V. Based on this consideration, we ignore the time for connection and disconnection of SMESS in the model. In addition, even if it may be hard to ignore this time due to reasons like deficient plug-and-play capability, we can also consider this time in the proposed strategy simply by adding it to the travel time $T_{j,ii'}$ and substituting the sum as the new travel time into the model.

Battery-based energy storage systems now play important roles in transmission and distribution systems. However, even though the use of MESS in bulk grids has been studied to prove its effectiveness, *e.g.*, the train-mounted MESS in [18], in our opinion, DS is still the main arena for MESS to show its remarkable skill, at least for now. Some reasons can be given as follows: Almost all existing MESSs on the market are mounted on trucks or trailers, having a capacity commonly lower than 1 MW·h [29]; and, in the scope of DS, these MESSs can be deployed within a reasonable time to match the operator's requirements. For example, in a DS, MESS can be deployed from the depot to a feeder or building that lost power due to disasters within acceptable hours; however, in a bulk grid that possibly covers multiple cities, a travel of MESS from its initial location to a bus of the grid may take days and a city in blackout can certainly not wait for that. In addition, when applied in DS, MESS can support different voltage-level use, *e.g.*, 0.38 – 35 kV in China [51]. Specifically, as mentioned before, MESS can be equipped with an onboard transformer, or even a separate mobile one, to match some uses of higher or lower voltage levels than designed for it [25], [29].

The interconnection of MESS, as well as other distributed energy resources, could affect the original protection function of the power grids. Adjustment of the protection settings is a necessary task that may take much effort, but we believe that the great advantages of MESS will make it worth the effort. Since the issue of protection settings is out of the scope of this paper, no further discussion is given herein.

## VII. Conclusion

Frequently occurring catastrophic events currently drive the requirement to enhance the power system survivability. In this paper, we propose a two-stage robust scheduling strategy to strengthen the IDS survivability by coordinating the two smart-grid technologies SMESSs and DR. With the survivability-oriented purpose, the SMESSs are scheduled to construct non-wires links reconnecting the external stranded sources and the IDS, which provide successive supplement of energy. Through alternate work, a continuous power supply can also be realized. DR is coordinated and scheduled to relieve the operating stress of the IDS in time. In addition to relieving power shortage, the relief to the energy shortage for IDS is also recognized to realize the proper energy use for the demand beyond the touch of FFGs in IDS under limited available energy. Numerical results show the effectiveness and advantages of the proposed strategy.

## References

[1] *Electric Power System Resiliency: Challenges and Opportunities*, document 3002007376, Electric Power Research Institute, Palo Alto, CA, USA, Feb. 2016.

[2] *Enhancing Distribution Resiliency: Opportunities for Applying Innovative Technologies*, document 1026889, Electric Power Research Institute, Palo Alto, CA, USA, Jan. 2013.

[3] W. Yuan *et al*, "Robust optimization-based resilient distribution network planning against natural disasters," *IEEE Trans. Smart Grid*, vol. 7, no. 6, pp. 2817-2826, Nov. 2016.

[4] S. Ma, S. Li, Z. Wang, and F. Qiu, "Resilience-oriented design of distribution systems," *IEEE Trans. Power Syst.*, vol. 34, no. 4, pp. 2880-2891, Jul. 2019.

[5] M. H. Amirioun, F. Aminifar, and H. Lesani, "Resilience-oriented proactive management of microgrids against windstorms," *IEEE Trans. Power Syst.*, vol. 33, no. 4, pp. 4275-4284, Jul. 2018.

[6] S. Lei, C. Chen, Y. Li, and Y. Hou, "Resilient disaster recovery logistics of distribution systems: co-optimize service restoration with repair crew and mobile power source dispatch," *IEEE Trans. Smart Grid*, vol. 10, no. 6, pp. 6187-6202, Nov. 2019.

[7] S. Yao, P. Wang, X. Liu, H. Zhang, and T. Zhao, "Rolling optimization of mobile energy storage fleets for resilient service restoration," *IEEE Trans. Smart Grid*, vol. 11, no. 2, pp. 1030-1043, Mar. 2020.

[8] C. Chen, J. Wang, F. Qiu, and D. Zhao, "Resilient distribution system by microgrids formation after natural disasters," *IEEE Trans. on Smart Grid*, vol. 7, no. 2, pp. 958-966, Mar. 2016.

[9] M. R. Kleinberg, K. Miu, and H. Chiang, "Improving service restoration of power distribution systems through load curtailment of in-service customers," *IEEE Trans. Power Syst.*, vol. 26, no. 3, pp. 1110-1117, Aug. 2011.

[10] C. Chen, J. Wang, and D. Ton, "Modernizing distribution system restoration to achieve grid resiliency against extreme weather events: an integrated solution," *Proc. IEEE*, vol. 105, no. 7, pp. 1267-1288, Jul. 2017.

[11] C. Keerthisinghe *et al*, "PV-battery systems for critical loads during emergencies: A case study from Puerto Rico after Hurricane Maria," *IEEE Power Energy Mag.*, vol. 17, no. 1, pp. 82-92, Jan.-Feb. 2019.

[12] R. Arghandeh, M. Pipattanasomporn, and S. Rahman, "Flywheel energy storage systems for ride-through applications in a facility microgrid," *IEEE Trans. Smart Grid*, vol. 3, no. 4, pp. 1955-1962, Dec. 2012.

[13] J. Nelson, N. G. Johnson, K. Fahy, and T. A. Hansen, "Statistical development of microgrid resilience during islanding operations," *Appl. Energy*, vol. 279, Dec. 2020, Art. no. 115724.

[14] A. Cattaneo, S. C. Madathil, and S. Backhaus, "Integration of optimal operational dispatch and controller determined dynamics for microgrid survivability," *Appl. Energy*, vol. 230, pp. 1685-1696, Nov. 2018.

[15] K. Balasubramaniam, P. Saraf, R. Hadidi, and E.B. Makram, "Energy management system for enhanced resiliency of microgrids during islanded operation," *Elect. Power Syst. Res.*, vol. 137, pp. 133-141, Aug. 2013.

[16] E. Rosales-Asensio, M. de Simón-Martín, D. Borge-Diez, J. J. Blanes-Peiŕo, and A. Colmenar-Santos, "Microgrids with energy storage systems as a means to increase power resilience: An application to office buildings," *Energy*, vol. 172, pp. 1005-1015, Apr. 2019.

[17] H. H. Abdeltawab and Y. A. I. Mohamed, "Mobile Energy Storage Scheduling and Operation in Active Distribution Systems," *IEEE Trans. Ind. Electron.*, vol. 64, no. 9, pp. 6828-6840, Sep. 2017.

[18] Y. Sun, Z. Li, M. Shahidehpour, and B. Ai, "Battery-based energy storage transportation for enhancing power system economics and security," *IEEE Trans. Smart Grid*, vol. 6, no. 5, pp. 2395-2402, Sep. 2015.

[19] J. Kim and Y. Dvorkin, "Enhancing distribution system resilience with mobile energy storage and microgrids," *IEEE Trans. Smart Grid*, vol. 10, no. 5, pp. 4996-5006, Sept. 2019.

[20] S. Lei, C. Chen, H. Zhou, and Y. Hou, "Routing and scheduling of mobile power sources for distribution system resilience enhancement," *IEEE Trans. Smart Grid*, vol. 10, no. 5, pp. 5650-5662, Sept. 2019.

[21] *One Solution for Every Unique Energy Need*, Nomad Transportable Power Systems LLC., Waterbury, VT, USA. Accessed: Sep. 28, 2021. [Online] Available: http://www.nomadpower.com/the-nomad-system/

[22] *TheBattery Mobile*, Alfen N.V., Almere, Netherlands. Accessed: Sep. 28, 2021. [Online] Available: http://alfen.com/sites/alfen.com/files/Datasheet-TheBatteryMobile-English.pdf

[23] *Y.Cube-Energy Storage: Product guide*. Aggreko Plc., Glasgow, UK. Accessed: Sep. 28, 2021. [Online]. Available: http://www.aggreko.com/-/media/Aggreko/Files/PDF/Energy-Storage/YCubeProductGuideFinal.pdf

[24] *Dynamic Power for A Dynamic Grid*. Power Edison LLC., Watchung, NJ, USA. Accessed: Sep. 28, 2021. [Online]. Available: https://www.poweredison.com/

[25] *Mobile Energy Storage*. Renewable Energy Systems Ltd., Hertfordshire, UK. Accessed: Sep. 28, 2021. [Online]. Available: http://www.res-group.com/media/342353/mobile_energystorage_28319.pdf

[26] Recent Projects. Greener Power Solutions B.V., Amsterdam, Netherlands. Accessed: Oct. 6, 2021. [Online]. Available: https://www.greener.nl/recent-projects/

[27] *Transportable and Mobile Energy Storage*, document 3002020153, Electric Power Research Institute, Palo Alto, CA, USA, Feb. 2021.

[28] Chapter 227 - An Act to Advance Clean Energy. The General Court of the Commonwealth of Massachusetts, Boston, MA, USA. Accessed: Oct. 6, 2021. [Online]. Available: https://malegislature.gov/Laws/SessionLaws/Acts/2018/Chapter227

[29] "Mobile energy storage study: Emergency response and demand reduction," Massachusetts Department of Energy Resources. Boston, MA, USA, Feb. 2020. [Online]. Available: http://www.mass.gov/doc/mobile-energy-storage-study/download

[30] W. Wang, X. Xiong, Y. He, J. Hu, and H. Chen, "Scheduling of separable mobile energy storage systems with mobile generators and fuel tankers to boost distribution system resilience," *IEEE Trans. Smart Grid*, early access.

[31] C. Gouveia, J. Moreira, C. L. Moreira, and J. A. Peças Lopes, "Coordinating storage and demand response for microgrid emergency operation," *IEEE Trans. Smart Grid*, vol. 4, no. 4, pp. 1898-1908, Dec. 2013.

[32] A. L. A. Syrri and P. Mancarella, "Reliability and risk assessment of post-contingency demand response in smart distribution networks," *Sustain. Energy, Grids and Netw.*, vol. 7, pp. 1-12, Sep. 2016.

[33] "NYISO 2019 annual report on demand response programs," New York Independent System Operator, Inc. Rensselaer, NY, USA, Jan. 2020.

[34] "Manual 7 - Emergency demand response program manual", New York Independent System Operator, Inc. (NYISO), Rensselaer, NY, USA, Apr. 2020.

[35] W. Wang, X. Xiong, C. Xiao, and B. Wei, "A novel mobility model to support the routing of mobile energy resources," 2020. [Online] Available: http://arxiv.org/abs/2007.11191

[36] N. Motegi, M. A. Piette, D. S. Watson, S. Kiliccote, and P. Xu, "Introduction to commercial building control strategies and techniques for demand response," Lawrence Berkeley National Laboratory, Berkeley, CA, USA, Rep. LBNL-59975, May 2007.

[37] M. E. Baran and F. F. Wu, "Network reconfiguration in distribution systems for loss reduction and load balancing," *IEEE Trans. Power Del.*, vol. 4, no. 2, pp. 1401-1407, Apr. 1989.

[38] J. Smith, A. Huskey, D. Jager, and J. Hur, "Wind turbine generator system safety and function test report for the Ventera VT10 wind turbine," National Renewable Energy Laboratory, Golden, Co, USA, Rep. NREL/TP-5000-56586, Nov. 2012.

[39] D. Bertsimas and M. Sim, "The price of robustness," *Oper. Res.*, vol. 52, no. 1, 35-53, Jan.-Feb. 2004.

[40] X. Chen, W. Wu, and B. Zhang, "Robust restoration method for active distribution networks," *IEEE Trans. Power Syst.*, vol. 31, no. 5, pp. 4005-4015, Sept. 2016.

[41] B. Zeng and L. Zhao, "Solving two-stage robust optimization problems using a column-and-constraint generation method," *Oper. Res. Lett.*, vol. 41, no. 5, pp. 457–461, 2013.

[42] H. Konno. "A cutting plane algorithm for solving bilinear programs," *Math. Program.*, vol. 11, pp. 14-27, 1976.

[43] V. Gabrel, M. Lacroix, C. Murat, and N. Remli. "Robust location transportation problems under uncertain demands," *Discrete Appl. Math.*, vol. 164, pp. 100-111, Feb. 2014.

[44] J. Löfberg, "YALMIP: A toolbox for modeling and optimization in MATLAB," in *Proc. IEEE Int. Symp. Comput. Aided Control Syst. Des.*, 2004, pp. 284–289

[45] Office of Energy Efficiency & Renewable Energy. Commercial and Residential Hourly Load Profiles for all TMY3 Locations in the United States. [Online]. Available: http://openei.org/datasets/dataset/commercial-and-residential-hourly-load-profiles-for-all-tmy3-locations-in-the-united-states

[46] California ISO. Today's outlook. [Online]. Available: http://www.caiso.com/TodaysOutlook/Pages/supply.html

[47] Caterpillar Inc. *200 ekW diesel generator set*. Accessed: Feb. 13, 2021. [Online]. Available: http://www.cat.com/en_US/products/new/power-systems/electric-power/diesel-generator-sets/106462.html

[48] W. Wang, X. Xiong, Y. He, and H. Chen. *Appendix for "Robust Survivability-Oriented Scheduling of Separable Mobile Energy Storage and Demand Response for Isolated Distribution Systems"*. [Online]. Available: http://drive.google.com/file/d/1Sl239unR822imyjv9xGhGLdOQCJ_WAEu/view?usp=sharing

[49] *IEEE Guide for Design, Operation, and Maintenance of Battery Energy Storage Systems, both Stationary and Mobile, and Applications Integrated with Electric Power Systems*, IEEE Standard 2030.2.1-2019, Dec. 2019.

[50] C. C. Cheng, J. T. Gao, and K. Y. Lo, "Hot-Swappable Grid-Connected Multilevel Converter for Battery Energy Storage System," *IEEE Trans. Circuits Syst. II, Express Briefs*, vol. 67, no. 10, pp. 2109-2113, Oct. 2020.

[51] *Technical requirements for mobile electrochemical energy storage system*, National Standard of the People's Republic of China GB/T 36545-2018, 2018.